\documentclass[a4paper, 12pt]{article}
\usepackage{epsfig,amssymb,euscript,xspace, color}
\usepackage{amsmath,jheppub,mathtools,rotating}
\def\bea{\begin{eqnarray}}
\def\eea{\end{eqnarray}}
\def\be{\begin{equation}}
\def\ee{\end{equation}}



\begin{document}

\title{
Holographic Conformal Partial Waves \\
~~~~~~~~~~ as Gravitational Open  Wilson Networks}
\author{Atanu Bhatta, Prashanth Raman and Nemani V. Suryanarayana}
\affiliation{Institute of Mathematical Sciences \\ C.I.T Campus, Taramani \\ Chennai 600113, India}
\emailAdd{batanu, prashanthr, nemani@imsc.res.in}

\abstract{
We propose a method to holographically compute the conformal partial waves in any decomposition of correlation functions of primary operators in conformal field theories using open Wilson network operators in the holographic gravitational dual. The Wilson operators are the gravitational ones where gravity is written as a gauge theory in the first order Hilbert-Palatini formalism. We apply this method to compute the global conformal blocks and partial waves in 2d CFTs reproducing many of the known results.}

\maketitle



\section{Introduction}
Gauge/gravity correspondence (holography) has been very helpful in studying quantum properties of both conformal field theories and gravitational theories. The original conjecture and subsequent works on AdS/CFT correspondence \cite{Maldacena:1997re, Witten:1998qj, Gubser:1998bc} provided a detailed prescription to compute the correlation functions of the CFT operators in terms of the so-called Witten diagrams in the bulk gravitational theory.

Most of the times the bulk side of the computation uses the metric (Einstein-Hilbert) formulation of gravity coupled to appropriate matter. In recent times there have emerged theories such as the higher spin gauge theories in three and higher dimensions involving gravity which are more naturally expressed as extensions of the tetrad (Hilbert-Palatini) formulation  than the metric formulation of the gravity sector. For example, the higher spin gauge theories \cite{Blencowe:1988gj} in three dimensions are described as Chern-Simon theories with the gauge algebra $sl(N, {\mathbb R}) \oplus sl(N, {\mathbb R})$ inside which the gravity sector in its Hilbert-Palatini formalism is embedded as an $sl(2, {\mathbb R}) \oplus sl(2, {\mathbb R})$ sector \cite{Achucarro:1987vz, Witten:1988hc}. Also it is well-known \cite{MacDowell:1977jt, Freidel:2005ak} that gravity with negative cosmological constant can be written as a BF-type gauge theory even in dimensions greater than three. It has been argued \cite{Doroud:2011xs} in the literature that Vasiliev's higher spin gauge theories \cite{Vasiliev:1989re, Vasiliev:1995dn} in four dimensions can be thought of as an appropriate extension of such a BF-type theory.  It appears to be more natural to describe these theories in this gauge theory formulation than in a metric-like formulation. They also are expected to describe appropriate conformal field theories holographically \cite{Gaberdiel:2012uj, Klebanov:2002ja}. Therefore it has become an important question to ask how to provide a prescription to compute the CFT quantities such as correlation functions. 

It is well known that the correlation function of a set of primary operators in a ($d$-dimensional Euclidean) CFT can be decomposed into its partial-waves. Thus the basic building blocks of the correlation functions of primary operators in a CFT are the partial waves which in-turn are made up of conformal blocks.\footnote{The conformal blocks of course are also an important input into the Bootstrap approach \cite{Ferrara:1973yt, Polyakov:1974gs, Mack:1975jr} to constrain the dynamics of any CFT.} For example the four-point function of four scalar primary operators can be decomposed as:
\bea
\langle {\cal O}_1(x_1) {\cal O}_2(x_2) {\cal O}_3(x_3) {\cal O}_4(x_4) \rangle = \sum_{{\cal O}} C_{12 {\cal O}} C^{\cal O}_{~~34} W_{\Delta, l} (x_i)
\eea
where $C_{12 {\cal O}}$ are the OPE coefficients and the partial wave $W_{\Delta, l} (x_i)$ is
\bea
W_{\Delta, l} (x_i) = \left(\tfrac{x_{24}^2}{x_{14}^2}\right)^{\tfrac{1}{2} (\Delta_1 - \Delta_2)} \left(\tfrac{x_{14}^2}{x_{13}^2}\right)^{\tfrac{1}{2} (\Delta_3 - \Delta_4)} \tfrac{G_{\Delta, l} (u,v)}{(x_{12}^2)^{\tfrac{1}{2}(\Delta_1+\Delta_2)} (x_{34}^2)^{\tfrac{1}{2}(\Delta_3+\Delta_4)} }
\eea
The pre-factor is determined by the conformal invariance and the function $G_{\Delta, l} (u,v)$ depends only on the conformally invariant cross ratios $u,v$. In general dimension (particularly in $d >2$ and when $d=2$ restricting to the large central charge $c$ limit) the partial wave $W_{\Delta, l} (x_i)$ is expected to satisfy two type of differential identities:
\bea
\label{globalwardidentity}
(M_{AB}^{(1)}+M_{AB}^{(2)} + M_{AB}^{(3)} + M_{AB}^{(2)}) W_{\Delta, l} (x_i) &=& 0 \cr
(M_{AB}^{(1)}+M_{AB}^{(2)})(M^{AB}_{(1)}+M^{AB}_{(2)}) W_{\Delta, l} (x_i) &=& C_2(\Delta, l) W_{\Delta, l} (x_i) \cr
&=& (M_{AB}^{(3)}+M_{AB}^{(4)})(M^{AB}_{(3)}+M^{AB}_{(4)}) W_{\Delta, l} (x_i)
\eea
where $M_{AB}^{(i)}$ is the operator representing the global conformal transformation generator acting on the primary operator ${\cal O}_i (x_i)$ and $C_2$ is the quadratic Casimir of the representation of the operator ${\cal O}$ being exchanged in the intermediate channel.   The first one is the reflection of the fact that the partial wave is covariant under the global conformal transformations. The second is the conformal Casimir equation \cite{Dolan:2000ut, Dolan:2003hv, Dolan:2011dv}.\footnote{If there are non-trivial higher order Casimir operators of the conformal algebra then one has to impose the corresponding differential equations as well.} These can be solved with appropriate boundary conditions to  obtain explicit expressions \cite{Dolan:2000ut, Dolan:2003hv, Dolan:2011dv} for the partial waves. The expression for $G_{\Delta, l} (u,v)$ in the partial wave of scalar operators are known in all even $d$ and in particular for $d=2$ (in the large $c$ limit) it is given by \cite{Dolan:2011dv}
\bea
\label{4ptscalarblock}
G_{\Delta, l} =  |z|^{\Delta-l}
\left[ z^l ~ {}_2F_1 \left( \tfrac{\Delta - \Delta_{12} +l}{2}, \tfrac{\Delta + \Delta_{34} +l}{2}, \Delta+l; z \right)  {}_2F_1 \left( \tfrac{\Delta - \Delta_{12} - l}{2}, \tfrac{\Delta + \Delta_{34} -l}{2}, \Delta-l; \bar z \right) + (z \leftrightarrow \bar z) \right] \nonumber
\eea
with $\Delta_{12} = \Delta_1 - \Delta_2$ etc., $z \bar z = \tfrac{x_{12}^2 x_{34}^2}{x_{13}^2 x_{24}^2} = u$ and $(1-z)(1- \bar z) = \tfrac{x_{14}^2 x_{23}^2}{x_{13}^2 x_{24}^2} = v$. One expects that the partial waves are also computable holographically. In fact the prescription to compute the conformal partial waves of the CFT in the language of Witten diagrams has been provided only recently in \cite{Hijano:2015zsa} which involves computing the so-called geodesic Witten diagrams. In this paper we will initiate addressing the question of how to compute partial waves (and conformal blocks) holographically in the gauge theory (Hilbert-Palatini) language of Euclidean $AdS_{d+1}$ gravity. 

In the Hilbert-Palatini formulation the gravitational fields, namely the vielbein 1-forms $e^a$ and the spin-connection 1-forms $\omega^{ab}$ can be packaged into one $so(1, d+1)$ adjoint valued gauge connection
\be
\label{gaugeconnection}
A = \frac{1}{2} \omega^{ab} M_{ab} + \frac{1}{l} e^a \, M_{0a}
\ee
where $M_{0a}$ and $M_{ab}$ are the generators of $so(1,d+1)$ with $a,b =1, \cdots d+1$. The parameter $l$ with dimensions of length sets the radius of $AdS_{d+1}$ vacuum. The action for the connection $A$ is in general an appropriate BF type theory \cite{MacDowell:1977jt, Freidel:2005ak}. The configurations satisfying
\be 
F = dA + A\wedge A = 0 
\ee
%
%
%
are the torsionless and locally $AdS_{d+1}$ spaces. One can couple matter particles as external sources to the $AdS_{d+1}$ gravity considering the Wilson line operators for the gauge connection (\ref{gaugeconnection}) in an appropriate representation of the gauge algebra along the curve given by the trajectory of the particle \cite{Alekseev:1988vx, Freidel:2006hv}.

The (global) partial waves constitute the non-dynamical building blocks of any CFT correlation function. Therefore one expects that they should be recovered holographically without putting too much information about the bulk theory and its interactions. Indeed, we propose that on the gravity side the expectation values of certain open Wilson network (OWN) operators for the connection (\ref{gaugeconnection}) that can be associated to directed trivalent spin-networks such as the one in Fig.(\ref{Fig1})  with their end points on the boundary of $AdS_{d+1}$ compute the partial waves of the dual CFT. 

For most part we will concentrate on $d=2$ case. Then the bulk contains Euclidean $AdS_3$ gravity described as the Chern-Simon's theory with gauge group $sl(2, {\mathbb R}) \oplus sl(2, {\mathbb R})$. In this case each edge of the spin network is labeled by an irreducible representation $(h, \bar h)$ of the gauge algebra $sl(2, {\mathbb R}) \oplus sl(2, {\mathbb R})$ and we associate to it an open Wilson line  in that representation connecting its end points. The representations of the external legs determine the dimensions of the operators of the CFT whose partial wave/conformal block we want to compute. We glue together the lines joining at a vertex an appropriate Clebsch-Gordan coefficient of the gauge algebra. We further project the external Wilson lines on to those states in the representation of that external leg that transform in a finite dimensional representation of the twisted diagonal sub-algebra of the gauge algebra $sl(2, {\mathbb R}) \oplus sl(2, {\mathbb R})$. Such states that transform in the trivial representation of the twisted diagonal $sl(2, {\mathbb R})$ have appeared in the recent literature in the construction of local bulk operators \cite{Miyaji:2015fia} (see also \cite{Nakayama:2015mva}).

\begin{center}
\setlength{\unitlength}{1.25cm}
\begin{picture}(8,8)(-1,-3)
\put(0,0){\vector(1,1){.5}}
\put(.5,.5){\line(1,1){.5}}
\put(1,1){\line(-1,1){1}}
\put(0,2){\vector(1,-1){.5}}
\put(1,1){\vector(2,0){.75}}
\put(1,1){\line(2,0){1.5}}
\put(2.5,1){\line(1,1){1}}
\put(2.5,1){\vector(1,1){.5}}
\put(2.5,1){\line(1,-1){1}}
\put(2.5,1){\vector(1,-1){.5}}
\put(3.5,2){\vector(0,1){.5}}
\put(3.5,2){\vector(1,0){.5}}
\put(3.5,2){\line(0,1){1}}
\put(5,2){\scalebox{0.7}{$|j_k, m_k \rangle\!\rangle$}}
\put(3.5,2){\line(1,0){1}}
\put(3.35,3.5){\begin{turn}{90}{\scalebox{0.7}{$|j_{n}, m_n \rangle\!\rangle$}}\end{turn}}
\put(3.4,-1.5){\begin{turn}{-90}{\scalebox{0.7}{$|j_i, m_i \rangle\!\rangle$}}\end{turn}}
\put(5,0){\scalebox{0.7}{$|j_j,m_j \rangle\!\rangle$}}
\put(3.5,0){\vector(1,0){.5}}
\put(3.5,0){\vector(0,-1){.5}}
\put(3.5,0){\line(1,0){1}}
\put(3.5,0){\line(0,-1){1}}
\put(4.6,2){$\dots$}
\put(3.45,3.1){$\vdots$}
\put(4.6,0){$\dots$}
\put(3.45,-1.45){$\vdots$}
\put(.2,1.85){\begin{turn}{-45}{\scalebox{0.7}{$(h_1, \bar h_1)$}}\end{turn}}
\put(.2,.05){\begin{turn}{45} {\scalebox{0.7}{$(h_2, \bar h_2)$}}\end{turn}}
\put(2.5,1.3){\begin{turn}{45} {\scalebox{0.7}{$(h', \bar h')$}}\end{turn}}
\put(2.5,0.6){\begin{turn}{-45} {\scalebox{0.7}{$(h'', \bar h'')$}}\end{turn}}
\put(1.3,1.1){\scalebox{0.7}{$(h, \bar h)$}}
\put(-.75,-.75){\begin{turn}{45}{\scalebox{0.7}{$\langle\!\langle j_2, m_2 |$}}\end{turn}}
\put(-.75,2.55){\begin{turn}{-45}{\scalebox{0.7}{$\langle\!\langle j_1, m_1 |$}}\end{turn}}
\put(0,-3){{\bf Fig. \!\!1}: A trivalent open spin network}
\label{Fig1}
\end{picture}
\end{center}
%
%
%
%
%
%
We will show that the classical limit of the vacuum expectation value of the open Wilson network operator corresponding to such a diagram obtained by simply evaluating it in the background of pure $AdS_3$ geometry satisfies the global conformal Ward identities such as (\ref{globalwardidentity}) when the external points are taken to the boundary. This is the bulk equivalent of the statement that the corresponding partial wave has to respect the global conformal symmetry of the CFT. We will also establish that they are solutions to the appropriate set of conformal Casimir equations such as the one in (\ref{globalwardidentity}). Then we evaluate such diagrams with two, three, four and five end points explicitly in the limit of the ends going to the boundary of the $AdS_3$ space and reproduce the known results.


The rest of the paper is organised as follows. In section 2 we elaborate on the construction of the open Wilson networks of interest and show that they satisfy the right identities such as the global conformal Ward identities and the conformal Casimir equations. In section 3 we initiate explicit evaluation of these diagrams and compare our results with the known answers in the literature. In section 4 we conclude with a discussion of the results and open questions. The appendices contain some background group theory and calculational details not presented in the text.

\section{Open Wilson Networks: definition and identities}
 We are interested in providing a prescription to compute partial waves of correlation functions of the dual $CFT_d$ in terms of the first order action of $AdS_{d+1}$ gravity. As alluded to in the introduction the basic ingredients are the gauge covariant and non-local Wilson line operators:
\be
W_y^x(R,C) = P \, {\rm exp} [{\int_y^x A}]
\ee
where $x$ and $y$ are two points in the space (with a boundary ${\mathbb R}^d$), $C$ is a curve connecting those, $R$ is a representation of the gauge algebra $so(1,d+1)$ and $A$ is the pull back of the gauge connection onto the curve $C$. As usual the symbol $P$ denotes the standard path ordering prescription. Under a gauge transformation $A \rightarrow h A h^{-1} + h dh^{-1}$ the Wilson line operator transforms covariantly as 
\be
W_y^x \rightarrow h(x) W_y^x h^{-1}(y) \, .
\ee
One can consider open Wilson network operators such as the one alluded to in the introduction where we take several directed open Wilson lines in different representations of the gauge algebra and glue the ends of any three lines ending at the same point by contracting the representation indices into the Clebsch-Gordan coefficients relating those representations. Such an open Wilson network operator will depend on the coordinates of the end points, the representations of each open Wilson line and, in general, the spin-network ${\cal N}$ that went into its construction. Let us call such an operator with $N$ end points $W_{\cal N}(x_1, R_1; x_2, R_2; \cdots; x_N, R_N)$. Then under gauge transformations it will transform covariantly as a tensor in the tensor product of all the representations $(R_1, \cdots, R_N)$. We will be interested in only those representations of  $so(1,d+1)$ which are related to the unitary (infinite dimensional) irreducible representation of the corresponding $so(2,d)$ relevant to the Lorentzian CFT.

In a general gauge theory we cannot hope that the expectation value of such an open Wilson network operator represents any physical quantity as it will not be gauge invariant. However the gauge transformations of (\ref{gaugeconnection}) with gauge group $SO(1,d+1)$ can be split into two subclasses representing both Local Lorentz transformations and the diffeomorphisms in the Euclidean $AdS_{d+1}$ gravity in the Hilbert-Palatini formulation. If we call the generators of $SO(1,d+1)$ as \{$M_{ab}$, $M_{0a}$\} with $a,b = 1, \cdots , d+1$ where $M_{ab}$'s generate the maximal compact subalgebra $so(d+1)$ and $M_{0a}$'s are like the boost operators of the Lorentz group - then the gauge transformation with parameter in the subalgebra $so(d+1)$ correspond to (the Euclidean analogs of) the Local Lorentz transformation of the vielbein $e^a$ and the spin-connection $\omega^{ab}$. 

%
Furthermore, since our theory is supposed to describe geometries that are asymptotically $AdS_{d+1}$ which have a boundary the observables  do not necessarily have to be invariant under all the gauge transformations but only under {\it small} gauge transformations, namely, those which do not have a non-trivial action on the boundary \cite{Brown:1986nw}.  

Now we are interested in computing the partial waves of a correlation function of a bunch of primary operators of the dual CFT. According to AdS/CFT the dual of a boundary primary operator is a bulk field. The fields in the bulk transforms in {\it finite} dimensional representations of the group of local Lorentz transformations. Therefore, we would first like to project the quantity $W_{\cal N}(x_1, R_1; x_2, R_2; \cdots; x_N, R_N)$ which is an element of the tensor product of the infinite dimensional representations $R_i$ of $so(1,d+1)$ down to that of the finite dimensional representations of the local Lorentz algebra $so(d+1)$. This step can be achieved by projecting the $i^{th}$ leg of the Wilson network operator in the representation $R_i$ of $so(1, d+1)$ onto vectors in this representation which provide the appropriate finite dimensional representation of the sub-algebra $so(d+1)$. As we will see such special states do exist and their construction is closely related to those in \cite{Miyaji:2015fia, Nakayama:2015mva}. It will turn out that one particular component of such a tensor has the leading fall off behaviour, as the points $x_i$ approach the boundary, compared to the other components. This component will be related to the partial wave of the corresponding primary operators of the CFT.

Having defined the open Wilson network (OWN) operators of interest classically, the next issue is how to define the expectation value of these operators in the quantum gauge theory. One can use a path integral definition \cite{Witten:1989wf} for this.  However we will not attempt to do this in this paper. Instead we will restrict ourselves to computing the values of these operators in the background of pure AdS space - which corresponds to evaluating the expectation values of these operators in the (semi-) classical limit. 

For any locally AdS space the corresponding gauge field strength of $A$ in (\ref{gaugeconnection}) vanishes. For such pure gauge configurations one can take $A = g \, dg^{-1}$ locally where $g$ is an element of $SO(1,d+1)$. Then it follows from the definition (\ref{gaugeconnection}) that such a configuration describes a given space with the corresponding $e^a$ and $\omega^{ab}$ satisfying the equation:
\be
\label{gequation}
dg + \frac{1}{2} \omega^{ab} M_{ab} \, g + \frac{1}{l} e^a M_{0a} \, g = 0.
\ee
If we are interested in finding the gauge field $A$ for a given space (with given $e^a$ and $\omega^{ab}$) we just have to solve this equation for $g$ and then use $A = - dg \, g^{-1}$.\footnote{The integrability condition of the equation (\ref{gequation}) reads:
\be
[\partial_\mu + \frac{1}{2} \omega_\mu^{ab} M_{ab}  +  \frac{1}{l} e_\mu^a M_{0a}, \partial_\nu + \frac{1}{2} \omega_\nu^{cd} M_{cd} +  \frac{1}{l} e_\nu^c M_{0c}] \, g(x) = 0
\ee
which may be written as:
\be
\frac{1}{2} [{R_{\mu\nu}}^{ab} + \frac{1}{l^2} (e^a_\mu e^b_\nu - e^a_\nu e^b_\mu)] M_{ab} \, g(x)  + \frac{1}{l} (\partial_\mu e^a_\nu - \partial_\nu e^a_\mu + \omega_\mu^{ab} e^b_\nu - \omega_\nu^{ab} e^b_\mu) M_{0a} \, g(x)  = 0
\ee
Thus any configuration that satisfies the equations $F = 0$ will lead to a $g$ and the integrability does not impose any further conditions. In higher dimensions integrability will impose non-trivial constraints as $F=0$ is not the equation on motion.} 
%
%
%
Notice that the equation (\ref{gequation}) for $g$ has a gauge invariance. It is covariant under an arbitrary local Lorentz transformation: $e^a \rightarrow (\Lambda)^{ac} e^c$, $\omega^{ab} \rightarrow (\Lambda)^{ac} \omega^{cd} (\Lambda^{-1})^{db} + (\Lambda)^{ac} d(\Lambda^{-1})^{cb}$ and $g \rightarrow \Lambda g$ where $\Lambda$ is any element of the subgroup $SO(d+1)$ and $(\Lambda)^{ab}$ are the matrix element of $\Lambda$ in the vector representation thus defining an equivalence relation between $g$ and $\Lambda g$. This makes the physical solution $g$ an element of the coset $so(1,d+1)/so(d+1)$. Notice also that the equation satisfied by $g$
%
%
is equivalent to 
\bea
\label{igequation}
dg^{-1} - \frac{1}{2} \omega^{ab} g^{-1} M_{ab} - \frac{1}{l} e^a g^{-1} M_{0a} = 0
\eea
This coset element $g$ turns out to be one of the ingredients in our prescription to compute boundary partial-waves. Before turning to the other ingredients let us  point out a relation between Killing vectors of the $AdS_{d+1}$ geometry and matrix elements of $g$ in the adjoint representation. It can be shown (see the appendix \ref{appendixb} for details) that the components of the Killing vectors $(l_{[\alpha\beta]})^\mu$ are given by
\bea
\label{kvfromadjg}
(l_{[\alpha\beta]})^\mu = - l \, E_a^\mu {(R[g^{-1}])_{\alpha\beta}}^{0a} = - l \, E_a^\mu {(R[g])^{0a}}_{\alpha\beta}
\eea
where $E^\mu_a$ is the inverse vielbein and ${(R[g^{-1}])_{\alpha\beta}}^{0a}$ are matrix elements of $g$ in its adjoint representation.\footnote{The reader familiar with Killing spinor equation of $AdS_{d+1}$ would probably recognise the similarity of it with the equations (\ref{gequation}, \ref{igequation}). The relation between the Killing vector components (\ref{kvfromadjg}) found here is a generalisation of a similar relation in the Killing spinor context to a more general representation of the local Lorentz algebra.}   We will make use of these facts to establish some differential equations satisfied by our OWN operators shortly. 

The second ingredient is the set of states in the representation space in which a given external Wilson line is in that transform in a (finite dimensional) irreducible representation of the sub algebra $so(d+1)$.\footnote{An identical problem has appeared in \cite{Nakayama:2015mva} in a closely related context.} We will construct several examples of such states later on in this paper particularly in the case of $d=2$ and make use of them to compute the OWN operators.

The last ingredient in the computation of the OWN expectation values is the Clebsch-Gordan (CG) coefficients of the gauge algebra $so(1,d+1)$. The expressions for these in the case of $d=2$ are derived in the appendix \ref{appendixA}.
\subsection{Partial waves of primary operators as OWN expectation value}
%
%
%
%
Now we are ready to provide the prescription to compute various boundary partial-waves in the classical limit. This can be obtained by simply evaluating the OWN in the flat connection corresponding to the $AdS_{d+1}$ geometry. Even though most of what we say in the rest of this section is applicable to higher dimensions we will have $d=2$ case in mind as the main illustrative example. The Wilson line evaluated in a flat connection (corresponding to a locally AdS geometry) is 
\bea
{\rm P} e^{ \int_{y}^x A} = g(x) g^{-1} (y)
\eea
This is taken in an irreducible (infinite dimensional) representation of $so(1,d+1)$ - particularly the one which would correspond to unitary representation of $so(2,d)$ which is the relevant gauge group of the Lorentzian case -- as mentioned earlier. Such a representation is labeled by the eigenvalues $(\Delta, l_1, \cdots, l_{[d/2]})$ of the Cartan generators of $so(1,d+ 1)$. On the other hand the representations of the subalgebra $so(d+1)$ are labeled by the ``angular momenta"  $(j_1, \cdots j_{[d/2]})$. We will label a state in the finite dimensional irrep of $so(d+1)$ found as a linear combination of states in the infinite dimensional irrep of $so(1,d+1)$ by $|\{\Delta, l_i\}: \{j_i\}, \{m_i\} \rangle\!\rangle$. 

%
%

We are now ready to form OWN operators that transform nicely under the local Lorentz rotation (LLR) algebra. We start with a spin-network of the type given in the introduction in Fig.(\ref{Fig1}). Associate to it a Wilson network operator as prescribed above. Then project each external leg with an outgoing arrow with a ket-type state in a representation of LLR algebra and each incoming external leg of the operator onto a bra-type dual state. This results in an object with $N$ floating indices (for an OWN with $N$ external legs) each of which transforms either by $R[h]$ or $R[h^{-1}]$ (depending on the index carried by the outgoing leg or the incoming leg of the OWN). It turns out that the quantity that satisfies the global conformal ward identities of a partial wave of the correlation function of primary operators corresponds to one particular component of this tensor. 

%
\subsubsection{Locations of vertices do not matter}

Since we are restricting ourselves to computing the expectation values of the OWN operators classically we simply evaluate them in the flat connection corresponding to the $AdS_{d+1}$ background. Now we will show that for this computation the positions of the vertices do not matter. For this we first note that at each vertex we have the following combination depending on the position of that vertex:\footnote{This is the relevant object when the trivalent vertex has two legs in representations $h_1$ and $h_2$ going out of it and the one in representation $h_3$ going into it. If the arrows are reversed on all legs then we simply have to replace the corresponding $g$ by $g^{-1}$.}
\bea
R_{h_1}[g(x)]_{k_1k_1'} ~ R_{h_2}[g(x)]_{k_2k_2'} ~ C^{h_1h_2;h_3}_{k_1'k_2';k_3'} ~R_{h_3}[g^{-1}(x)]_{k_3'k_3} 
\eea
where $R_h[g(x)]$ is the matrix representation of the coset element $g(x)$ that we have defined earlier in (\ref{gequation}) and $C^{h_1h_2;h_3}_{k_1'k_2';k_3'}$ is the relevant CG coefficient. Now we use the identity (\ref{noxatvertex}) in appendix \ref{appendixA} to replace this by one CG coefficient eliminating the coordinate dependence of the junction. This can be done at every trivalent vertex in our spin-network thus eliminating the dependence of the locations of the vertices as claimed.

\subsubsection{Differential equations satisfied}
The global blocks/partial waves are expected to satisfy some differential relations as stated in the introduction. Now we want to show that an OWN such as the one in Fig.(\ref{Fig1}) will also satisfy the same set of differential identities expected of  the corresponding partial wave. To proceed further we note the following identities:
\bea
\label{gmmig}
g(x) \, M_{\alpha\beta} &=& l^\mu_{\alpha\beta} (x) \partial_\mu g(x) + \frac{1}{2} M_{bc} g(x) \left[ 
\omega_\mu^{bc} (x) l^\mu_{\alpha\beta} (x) + {(R[g(x)])^{bc}}_{\alpha\beta} \right] \cr
 M_{\alpha\beta} g^{-1} (x) &=& -l^\mu_{\alpha\beta} (x) \partial_\mu g^{-1} (x) + \frac{1}{2} \left[ 
\omega_\mu^{bc} (x) l^\mu_{\alpha\beta} (x) + {(R[g(x)])^{bc}}_{\alpha\beta} \right]  g^{-1} (x) M_{bc}
\eea
where $g(x)$ ($g^{-1}(x)$) is a solution to (\ref{gequation}) ((\ref{igequation})), $l^\mu_{\alpha\beta} (x)$ are the components of the Killing vector of the background geometry carrying the indices of the corresponding algebra generator $M_{\alpha\beta}$ of the left hand side. It can be verified that these Killing vectors $l^\mu_{\alpha\beta} (x) \partial_\mu$ satisfy the same algebra as their corresponding algebra generators $M_{\alpha\beta}$. 

The ingredients in our OWN are the matrix elements of $g(x)$ or $g^{-1}(x)$ between a generic state in the representation of the particular external leg and the state in the finite dimensional representation of the sub algebra $so(d+1)$. It turns out that these quantities in the limit of bulk point $x$ approaching the boundary of $AdS_{d+1}$ can be computed (will be done explicitly for $d=2$ case in the next section). We can also compute these with either the additional insertions of $M_{\alpha\beta}$  to the right of $g(x)$ or $-M_{\alpha\beta}$ to the left of $g^{-1}(x)$ depending on the direction of the external leg. It can be shown (again will be exhibited explicitly for the $d=2$ case in the next section) further that these matrix elements simply turn out to be those obtained by the action of the boundary conformal transformation of a primary (or descendent) operator on the matrix element without the insertion in the boundary limit. Now the left hand side of the global conformal Ward identity is simply given by the boundary limit of sum of the OWN operators with the insertion of the corresponding gauge generator ($M_{\alpha\beta}$ after $g(x)$ if it is an ingoing leg and $-M_{\alpha\beta}$ before $g^{-1}$ for the outgoing one). Using the recursion relations that the CG coefficients are expected to satisfy it can be seen that this sum will vanish. In the particular case of $d=2$ we will demonstrate this fact shortly. This establishes the identity that under simultaneous transformation of the primary operators under the global conformal transformations the OWN expectation value is left invariant.

Now we turn to the Casimir equations that the global conformal blocks are expected to satisfy. Because the partial wave decomposition of a correlation function involves taking the contribution of one primary (and its global descendants) they are expected to satisfy the conformal Casimir equation with eigenvalue given by the Casimir invariant of the primary in question. One expects one Casimir equation for each channel of decomposition of the correlator. In our context this translates to the expectation that our OWN operator (associated to a spin-network such as the one in Fig.(\ref{Fig1})) satisfies a Casimir equation for each (``1-particle reducible'') edge of the spin network graph that when cut the diagram falls apart into two disjoint pieces (which is the case of any intermediate leg of a tree-level network, i.e., without closed loops). We now want to argue that this is indeed the case for our diagrams. 

We will use the 4-point partial wave (in the $s$-channel decomposition) in $d=2$ as the illustrative example. In this case there are two independent quadratic Casimir operators (one for each of the two commuting $sl(2, {\mathbb R})$ algebras in $so(1,3)$). The partial wave is expected to satisfy one Casimir equation corresponding to the quadratic Casimir of the full algebra $so(1,3)$. However our OWNs satisfy two equations - one for each of the two quadratic Casimirs of $so(1,3)$. It will turn out that there are two OWNs for each intermediate (``1-particle reducible'') edge (connecting two trivalent vertices) with the same eigenvalue of the quadratic Casimir of the full algebra $so(1,3)$ related by the interchange $h \leftrightarrow \bar h$ in that edge. Therefore any linear combination of these two OWNs will satisfy the Casimir equation. Then one should be guided by the boundary conditions expected (from the OPEs in the CFT as one takes the coincidence limits of various vertices). We will comment on this aspect again later on.

The value of the 4-point network (see Fig. \!\!4) corresponding to the partial wave of 4-point function of primary operators is \footnote{The particular projection on to specific states in the finite dimensional representation of $so(3)$ chosen below will be justified in the next section.} 
\bea
W_{h, \bar h}(x_i) &=& \langle\!\langle h_1, \bar h_1; j_1, j_1 | g(x_1) |h_1, \bar h_1; k_1, \bar k_1 \rangle
\langle\!\langle h_2, \bar h_2; j_2, j_2 | g(x_2) |h_2, \bar h_2; k_2, \bar k_2 \rangle \cr
&& \langle h_3, \bar h_3; k_3, \bar k_3| g^{-1}(x_3) | h_3, \bar h_3; j_3, -j_3 \rangle\! \rangle
\langle h_4, \bar h_4; k_4, \bar k_4| g^{-1}(x_4) | h_4, \bar h_4; j_4, -j_4 \rangle\! \rangle \cr
&& \times C^{h_1h_2h}_{k_1, k_2,k} \times C^{\bar h_1 \bar h_2 \bar h}_{\bar k_1 \bar k_2 \bar k} \times C^{h_3 h_4 h}_{k_3 k_4 k} \times C^{\bar h_3 \bar h_4 \bar h}_{\bar k_3 \bar k_4 \bar k}
\eea
where the sum over repeated indices is assumed. Then the action of the Casimir differential operator on the partial wave in eq.(\ref{globalwardidentity})  
\bea
(M_{AB}^{(1)}+M_{AB}^{(2)})(M^{AB}_{(1)}+M^{AB}_{(2)}) W_{h, \bar h} (x_i)
\eea
is obtained by summing over three diagrams with the first one with an insertion of the Casimir operator $M_{\alpha \beta} M^{\alpha\beta}$ after $g(x_1)$, the second one with an insertion of $ M_{\alpha \beta} M^{\alpha\beta}$ after $g(x_2)$ and the third one with one insertion of $M_{\alpha\beta}$ after $g(x_1)$ and one insertion of $M^{\alpha\beta}$ after $g(x_2)$ with a factor of two and summing over the $\alpha$ and $\beta$ indices. Let us consider the Casimir made of $\{L_0, L_{\pm1}\}$ first  (see appendix \ref{appendixA} for our conventions on $so(1,3)$ generators). Then the answer of this sum contains the following terms:
\bea
&& (L_a^{(1)} L^a_{(1)})_{k_1 k_1'} C^{h_1 h_2 h}_{k_1' k_2 k} + (L_a^{(2)} L^a_{(2)})_{k_2k_2'} C^{h_1 h_2 h}_{k_1 k_2' k} + (L_a^{(1)})_{k_1k_1'}  (L^a_{(2)})_{k_2k_2'} C^{h_1 h_2 h}_{k_1' k_2' k} + (L^a_{(1)})_{k_1 k_1'}  (L_a^{(2)})_{k_2 k_2'} C^{h_1 h_2 h}_{k_1' k_2' k}  \cr 
&&= (L^a_{(1)})_{k_1 k_1''} \left[ (L_a^{(1)})_{k_1''k_1'} C^{h_1 h_2 h}_{k_1' k_2 k} + (L_a^{(2)})_{k_2 k_2'} C^{h_1 h_2 h}_{k_1'' k_2' k} \right] + (L^a_{(2)})_{k_2k_2''} \left[ (L_a^{(1)})_{k_1 k_1'} C^{h_1 h_2 h}_{k_1' k_2'' k} + (L_a^{(2)})_{k_2'' k_2'} C^{h_1 h_2 h}_{k_1 k_2' k} \right] \cr
&&=  (L^a_{(1)})_{k_1 k_1'}  C^{h_1 h_2 h}_{k_1' k_2 k'} (L_a^{(0)})_{k'k}+ (L_a^{(2)})_{k_2 k_2'}  C^{h_1 h_2 h}_{k_1 k_2' k'} (L_a^{(0)})_{k'k} \cr
&&= \left[ (L^a_{(1)})_{k_1 k_1'}  C^{h_1 h_2 h}_{k_1' k_2 k'} + (L_a^{(2)})_{k_2 k_2'}  C^{h_1 h_2 h}_{k_1 k_2' k'} \right] (L_a^{(0)})_{k'k} = C^{h_1 h_2 h}_{k_1 k_2 k''} (L^a_{(0)})_{k'' k'}(L_a^{(0)})_{k'k}
\eea
where we denoted the matrix elements of the generator $L_a$ in the representation $(h_1, \bar h_1)$ by $L_a^{(1)}$ etc. and in representation $(h, \bar h)$ by $L_a^{(0)}$. Substituting this result back into the sum of diagrams we started with we see that the result is simply given by the value of the Casimir operators $L_aL^a$ in the representation $(h, \bar h)$ times the original diagram. It can be easily checked that our digram satisfies the corresponding Casimir equation for the second quadratic Casimir operator made of $\{\bar L_0, \bar L_{\pm1}\}$ as well. This proof generalises to any spin-network straightforwardly. The identity can also be generalised to higher dimensions as well - though we will not consider computing this case in this paper. Finally one just needs to ensure that the right boundary conditions are imposed to show that our OWN operators indeed compute the partial waves. 

Having defined and elaborated on the OWN operators we now turn to computing them explicitly in the $d=2$ case. In the course of this various properties argued for in this section will be demonstrated.

\section{Euclidean $AdS_3$ with boundary ${\mathbb R}^2$}
 We consider the Eucidean $AdS_3$ geometry with boundary ${\mathbb R}^2$. The metric is
\bea
l^{-2} \, ds^2 = d\rho^2 + e^{2\rho} (dx_1^2 + dx_2^2)
\eea
where $l$ is the radius of $AdS_3$ and the ranges of the coordinates are $-\infty < \rho, x_1, x_2 < \infty$ as usual. In these coordinates $\rho \rightarrow \infty$ is the conformal boundary. The Killing vectors of this geometry are:
\bea
L_{-1} &=& - \partial_z, ~~ L_0 = \tfrac{1}{2} \partial_\rho - z \partial_z, ~~ L_1 = z \partial_\rho - z^2 \partial_z + e^{-2\rho} \partial_{\bar z} \cr
\bar L_{-1} &=& - \partial_{\bar z}, ~~ \bar L_0 = \tfrac{1}{2} \partial_\rho - {\bar z} \partial_{\bar z}, ~~ \bar L_1 = {\bar z} \partial_\rho - {\bar z}^2 \partial_{\bar z} + e^{-2\rho} \partial_z 
\eea
which satisfy the commutator algebra $[L_m, L_n] = (m-n) L_{m+n}, [\bar L_m, \bar L_n] = (m-n) \bar L_{m+n}, [L_m, \bar L_n] = 0$. Let us choose the frame: $e^1 = l \, e^\rho \, dx_1$, $e^2 = l \, e^\rho \, dx_2$, $e^3= l \, d\rho$. Then the non-vanishing spin-connections are: $\omega^{a3} = \frac{1}{l} e^a$ for $a=1,2$. The equation $dg + \frac{1}{2} \omega^{ab} M_{ab} g + \frac{1}{l} e^a M_{0a} g = 0$ satisfied by the coset element $g$ in this case reads
\bea
dg + d\rho \,  M_{03} g + e^\rho dx_1 \, (M_{13}+M_{01}) g   + e^\rho dx_2 \,  (M_{23} + M_{02})   g  = 0.
\eea
Its solution may be written as
\bea
\label{gone}
g = e^{-\rho M_{03}} e^{- x_1 (M_{13}+M_{01}) } e^{- x_2 (M_{23}+M_{02}) }
\eea
%
%
%
%
%
%
up to a multiplication by a constant group element on the right.\footnote{This solution generalises easily to $AdS_{d+1}$ geometry with ${\mathbb R}^d$ boundary.} Written in terms of the generators of the two $sl(2, {\mathbb R})$ factors in $so(1,3)$ this reads:
\bea
\label{gtwo}
g = e^{\rho (L_0 + \bar L_0)} e^{-L_{-1} (x_1+i x_2)}  e^{-\bar L_{-1} (x_1 -i x_2)} \, .
\eea
%
%
%
The basic building blocks of the open Wilson networks are again $\langle h, \bar h; k, \bar k| g^{-1}(x) |h, \bar h; j, m \rangle \!\rangle $ and $\langle\!\langle h, \bar h; j, m  | g(x) | h, \bar h; k, \bar k \rangle$. To evaluate these we need the states $|h, \bar h; j, m \rangle \!\rangle$ and $\langle\!\langle h, \bar h; j, m  |$.
\subsection{Constructing the states $|h, \bar h; j,m \rangle\!\rangle$}
We define $|h, \bar h; j, m \rangle \!\rangle $ as follows:\footnote{When the bulk field dual to the primary operator at hand is a gauge field one expects classes of the cap states reflecting the gauge symmetry \cite{Nakayama:2015mva} -- we do not consider this case here. } 
\bea
|h, \bar h; j, m\rangle\!\rangle = \lambda (h, \bar h) \sum_{n=0}^\infty (-1)^{\frac{p}{2}+n} \sqrt{\tfrac{(n+p)!}{n!p!}} \sqrt{\tfrac{\Gamma(2\bar h +n)}{\Gamma(2h+n+p) \Gamma(2\bar h - 2h +1 - p)}} ~~ |h, n+p\rangle \otimes |\bar h, n \rangle
\eea
where $p = j+m$ and $\bar h - h = j \ge 0$.\footnote{When $h-\bar h \ge 0$ one can write a similar set of states obtained by interchanging $h$ with $\bar h$ and the order of the  states in the tensor product.} We assume that $j\in {\mathbb Z}/2$ and thus $p=0, 1, \cdots 2j$. The factor $\lambda(h , \bar h)$ is arbitrary constant at this stage - will be chosen to be 
\bea
\label{lambda}
\lambda(h , \bar h)^2 = (-1)^{\bar h - h} \tfrac{(2\bar h - 2h)! \Gamma(2h)}{\Gamma(2\bar h)} 
\eea
for convenience. Notice that when $j=0$ this state is closely related to the one written down in \cite{Miyaji:2015fia} and for other $j$ advocated for in \cite{Nakayama:2015mva}. These states can be shown to satisfy 
\bea
(L_0 - \bar L_0) |h, \bar h; -j, p\rangle\!\rangle &=& (-j+p) |h, \bar h; -j, p\rangle\!\rangle \cr
(L_{-1} + \bar L_1) |h, \bar h; - j, p\rangle\!\rangle &=& \sqrt{(p+1) (-2j+p)} ~  |h, \bar h; -j, p +1 \rangle\!\rangle \cr
(L_1 + \bar L_{-1}) |h, \bar h; - j, p\rangle\!\rangle &=& \sqrt{p (-2j+p -1)} ~  |h, \bar h; -j, p -1 \rangle\!\rangle
\eea
corresponding to $h^D = -j$ as expected from \cite{Jackiw:1990ka} and therefore form a non-unitary finite dimensional representation of the twisted diagonal $sl(2,{\mathbb R})$ generated by $\{L_0^D := L_0 - \bar L_0, L_1^D := L_1 + \bar L_{-1}, L_{-1}^D := L_{-1} + \bar L_1\}$. The local Lorentz group $SO(3)$ is generated by $\{J_3 = L_0^D, J_+ = \pm i L_{-1}^D, J_- = \pm i L_1^D\}$. Then these states can be seen to provide the unitary representation labeled by the angular momentum $j$ (with the identification $|h, \bar h; -j, p\rangle\!\rangle \rightarrow |h, \bar h; j, m\rangle\!\rangle$ which we will use interchangeably) of the $su(2)$ algebra generated by these $\{J_3, J_{\pm }\}$.
\subsection{Computing $\langle\!\langle h, \bar h; j, m  | g(x) | h, \bar h; k, \bar k \rangle$ in $\rho\rightarrow \infty$ limit}
We start with eq.(\ref{gtwo}) with $z = x_1 + i x_2$
\bea
g(x) = e^{\rho (L_0 + \bar L_0) }e^{- z L_{-1} } e^{-\bar z \bar L_{-1}}
\eea
Then 
\bea
&&\langle\!\langle h, \bar h; -j, p  | g(x) | h, \bar h; k, \bar k \rangle \cr
&=& \lambda \sum_{n=0}^\infty (-1)^{\frac{p}{2}+n} \sqrt{\tfrac{(n+p)!}{n!p!}} \sqrt{\tfrac{\Gamma(2\bar h +n)}{\Gamma(2h+n+p) \Gamma(2\bar h - 2h +1 - p)}}  \cr
&& \times \langle h, n+p| e^{\rho L_0} e^{-z L_{-1}} |h, k\rangle \langle \bar h, n | e^{\rho \bar L_0} e^{-\bar z \bar L_{-1}} |\bar h, \bar k\rangle \cr
&=& \lambda  \tfrac{(-z)^{p-k} (-\bar z)^{-k} e^{\rho (h + \bar h +p)} (-1)^{n+\frac{p}{2}}}{\sqrt{\Gamma(2h+k) \Gamma(2\bar h + \bar k) k! \bar k! p! \Gamma(2j+1-p)}}\sum_{n={\rm max}(k-p, \bar k)}^\infty \tfrac{(n+p)! \Gamma(2\bar h +n)}{(n+p-k)! (n-\bar k)!} \left(-e^{2\rho} |z|^2 \right)^n \cr
&{k-p \ge \bar k \atop =}&  \lambda  \tfrac{e^{\rho (h + \bar h +p)} (- z)^{p-k} (-\bar z)^{-\bar k} k! \Gamma(2\bar h + k - p) (-1)^{p/2}}{\sqrt{k! \bar k! p! \Gamma(2h+k) \Gamma(2 \bar h+\bar k) \Gamma(2\bar h - 2h +1 -p) }} \tfrac{(-e^{2\rho} |z|^2)^{k-p}}{(k - \bar k -p)!} \cr
&& ~~ ~~{} _2F_1 (k+1, 2 \bar h+k-p, k-p-\bar k +1, - e^{2\rho} |z|^2) \cr
&& \cr
&{k-p \le \bar k \atop =}&  \lambda  \tfrac{e^{\rho (h + \bar h +p)} (- z)^{p-k} (-\bar z)^{-\bar k} \Gamma(2\bar h + \bar k) (\bar k +p)! (-1)^{p/2}}{\sqrt{p! k! \bar k! \Gamma(2h+k) \Gamma(2 \bar h+\bar k) \Gamma(2\bar h - 2h +1 -p)}} \tfrac{(-e^{2\rho} |z|^2)^{\bar k}}{(\bar k - k +p)!} \cr 
&& ~~~~ \times {}_2F_1 (\bar k+1+p, 2 \bar h+ \bar k, p+\bar k- k +1, - e^{2\rho} |z|^2) \nonumber 
\eea
We would like to take the $\rho \rightarrow \infty$ of these expressions. For this we use the well-known Euler's identity (see, for instance, \cite{aar1999})
\bea
{}_2F_1(\alpha, \beta, \gamma, x) = (1-x)^{\gamma-\alpha-\beta} {}_2F_1(\gamma- \alpha, \gamma -\beta, \gamma, x)
\eea
Using this we have
\bea
{}_2F_1 (k+1, 2 \bar h+k-p, k-p-\bar k +1, - e^{2\rho} |z|^2) &{ \rho \rightarrow \infty \atop \longrightarrow}&  (-1)^{p+\bar k} (e^{2\rho} |z|^2)^{p-k-2\bar h} \tfrac{k! (p+k)! \Gamma(2 \bar h + \bar k)}{(k-\bar k - p)! \Gamma(2\bar h - p)} \cr 
{}_2F_1 (\bar k+1+p, 2 \bar h+ \bar k, p+\bar k- k +1, - e^{2\rho} |z|^2) &{ \rho \rightarrow \infty \atop \longrightarrow}& (-1)^k (e^{2\rho} |z|^2)^{-\bar k-2\bar h} \tfrac{ (\bar k- k + p)! \Gamma(2 \bar h + k -p)}{\Gamma(2\bar h - p)(p+\bar k)!} \cr && 
\eea
When we take the $\rho \rightarrow \infty$ limit both the cases reduce to the same expression given by
\bea
\langle\!\langle h, \bar h; -j, p  | g(x) | h, \bar h; k, \bar k \rangle
  \rightarrow \lambda  (-1)^{-p/2} \tfrac{ e^{\rho (h-3\bar h +p)} z^{-2\bar h} {\bar z}^{-2\bar h} z^{p-k} \bar z^{-\bar k}}{\sqrt{k! \bar k! \Gamma(2h+k) \Gamma(2\bar h + \bar k) \Gamma(2j +1 -p)p!}} \tfrac{\Gamma(2\bar h + \bar k)\Gamma(2\bar h - p+k)}{\Gamma(2\bar h -p)} + \cdots \nonumber
\eea
At this point let us note that since $p$ runs from $0$ to $2j$ the leading terms in the $\rho \rightarrow \infty$ limit comes by setting $p = 2j$ which goes as $e^{-\rho (h + \bar h)}$ and lower values of $p$ lead to sub-leading terms in this limit. Therefore, the special case of $p=2j$ should correspond to insertion of a primary operator at the corresponding boundary point. To substantiate this we now show that the matrix element $\langle\!\langle h, \bar h; j, m  | g(x) M_{\alpha\beta} | h, \bar h; k, \bar k \rangle$ is the conformal transformation of the answer without the insertion of $M_{\alpha\beta}$. By explicit computation of eqs.(\ref{gmmig}) we find
\bea
g(x) L_{-1} &=& -\partial_z g(x) \cr
g(x) L_0 &=& -z \partial_z g(x) + \frac{1}{2} (\partial_\rho + (L_0 - \bar L_0)) g(x) = (-z \partial_z + L_0) g(x) \cr
g(x) L_1 &=& -z^2 \partial_z g(x) +  z (\partial_\rho +(L_0 - \bar L_0)) g(x)  + e^{-\rho} (L_1 + \bar L_{-1}) g(x) + e^{-2\rho} \partial_{\bar z} g(x) \cr
g(x) \bar L_{-1} &=& - \partial_{\bar z} g(x) \cr
g(x) \bar L_0 &=& - \bar z \partial_{\bar z} g(x) + \frac{1}{2} (\partial_\rho - (L_0 - \bar L_0)) g(x)  \cr
g(x) \bar L_1 &=&  - \bar z^2 \partial_{\bar z} g(x) + {\bar z} (\partial_\rho - (L_0 - \bar L_0)) g(x) + e^{-\rho} (L_{-1} + \bar L_1) g(x) + e^{-2\rho} \partial_{z} g(x)
\eea
Notice that the terms leading in $\rho \rightarrow \infty$ limit do not mix different states in the irrep of the twisted diagonal $sl(2, {\mathbb R})$ whereas the sub-leading ones do. Using matrix elements just computed we can write down the effect of insertion of $-L_n$'s and $-\bar L_n$'s. We find:
\bea
-\langle\!\langle h, \bar h; j, j  |g(x) L_{-1} | h, \bar h; k, \bar k \rangle &=& \partial_z \langle\!\langle h, \bar h; j, j  |g(x) | h, \bar h; k, \bar k \rangle \cr
-\langle\!\langle h, \bar h; j, j  |g(x) L_0 | h, \bar h; k, \bar k \rangle&=& (z \partial_z + h) \langle\!\langle h, \bar h; j, j  | g(x) | h, \bar h; k, \bar k \rangle \cr
-\langle\!\langle h, \bar h; j, j  |g(x) L_1 | h, \bar h; k, \bar k \rangle&=& (z^2 \partial_z + 2 z h)  \langle\!\langle h, \bar h; j, j  | g(x) | h, \bar h; k, \bar k \rangle \cr
-\langle\!\langle h, \bar h; j, j  |g(x) \bar L_{-1} | h, \bar h; k, \bar k \rangle&=&  \partial_{\bar z} \langle\!\langle h, \bar h; j, j  | g(x) | h, \bar h; k, \bar k \rangle \cr
-\langle\!\langle h, \bar h; j, j  |g(x) \bar L_0 | h, \bar h; k, \bar k \rangle&=&  (\bar z \partial_{\bar z} + \bar h)  \langle\!\langle h, \bar h; j, j  | g(x) | h, \bar h; k, \bar k \rangle  \cr
-\langle\!\langle h, \bar h; j, j  |g(x) \bar L_1 | h, \bar h; k, \bar k \rangle&=&   (\bar z^2 \partial_{\bar z} + 2 \bar z \bar h) \langle\!\langle h, \bar h; j, j  |g(x) | h, \bar h; k, \bar k \rangle
\eea
in the $\rho \rightarrow \infty$ limit -- establishing their transformation properties as primaries under the corresponding conformal transformation (see for instance \cite{DiFrancesco:1997nk}). What this means is that the functions $\langle\!\langle h, \bar h; j, j  |g(x) | h, \bar h; k, \bar k \rangle$ in the $\rho \rightarrow \infty$ limit provide a representation of the conformal algebra $sl(2, {\mathbb R}) \oplus sl(2, {\mathbb R})$ -- the first factor acting on $z$ and the second factor acting on $\bar z$. Remarkably the representation (for either the holomorphic or the antiholomorphic part) that we find here is the same  as the one used in \cite{Bargmann:1946me} (see also \cite{Jackiw:1990ka}) in the study of unitary irreps of $sl(2, {\mathbb R})$ algebra around $z\rightarrow \infty$ in the complex plane.

\subsection{Computing $\langle h, \bar h; k, \bar k| g^{-1}(x) |h, \bar h; j, m \rangle \!\rangle $ in $\rho\rightarrow \infty$ limit}
This can also be computed and the calculation is simpler than the previous one. Starting with 
\bea
g^{-1}(x) = e^{\bar z \bar L_{-1}} e^{z L_{-1}} e^{-\rho (L_0 + \bar L_0)}
\eea
we obtain
\bea
&& \langle h, \bar h; k, \bar k| g^{-1}(x) |h, \bar h; j, m \rangle \!\rangle \cr
&=&  \lambda  \sum_{n=0}^\infty (-1)^{n+\frac{p}{2}} \sqrt{\tfrac{(n+p)!}{n!p!}} \sqrt{\tfrac{\Gamma(2\bar h +n)}{\Gamma(2h+n+p) \Gamma(2\bar h - 2h +1 - p)}}  \langle h, k| e^{z L_{-1}} e^{-\rho L_0} | h, n+p \rangle \langle \bar h, \bar k| e^{\bar z \bar L_{-1}} e^{-\rho \bar L_0} | \bar h, n \rangle \cr
&=& \lambda  (-1)^{p/2} e^{-\rho (h + \bar h + p)} \sqrt{\tfrac{\Gamma(2h+k) \Gamma(2 \bar h+\bar k) k! \bar k!} {p! \Gamma(2\bar h - 2h +1 - p)}} \sum_{n=0}^{{\rm min}(k-p, \bar k)} \tfrac{(-e^{-2\rho} |z|^{-2})^n}{n! (\bar k - n)! \Gamma(2h+n+p) (k-n-p)!} \cr
&=& \lambda  (-1)^{p/2} \tfrac{e^{-\rho (h+\bar h+p)}}{\Gamma(2h+p)}  \sqrt{\tfrac{k! \bar k! \Gamma(2h+k) \Gamma(2 \bar h+\bar k)} {p! \Gamma(2\bar h - 2h +1 - p)}}\tfrac{ z^{k-p}}{(k-p)!} \tfrac{ \bar z^{\bar k}}{\bar k!}   ~~ {}_2F_1[-\bar k, -k+p, 2h+p; - e^{-2\rho} |z|^{-2}] \cr
& {\rho \rightarrow \infty \atop =} & \lambda  (-1)^{p/2} \tfrac{e^{-\rho (h+\bar h+p)}}{\Gamma(2h+p)}  \tfrac{ z^{k-p}}{(k-p)!} \tfrac{ \bar z^{\bar k}}{\bar k!}   \sqrt{\tfrac{k! \bar k! \Gamma(2h+k) \Gamma(2 \bar h+\bar k)} {p! \Gamma(2\bar h - 2h +1 - p)}}+ {\cal O}(e^{-\rho (h+\bar h +p+1)})
\eea
Let us note that the leading term from this leg comes from $p=0$ ($m=-j$) which again goes as $e^{-\rho (h+\bar h)}$ and higher values of $p$ give sub-leading terms in the $\rho \rightarrow \infty$ limit. In this case the $p=0$ answer corresponds to insertion of a primary with dimensions $(h, \bar h)$ at the boundary point. This can again be seen on similar lines as before by first observing the identities (\ref{gmmig}):
\bea
L_{-1}  g^{-1}(x) &=& \partial_z g^{-1}(x) \cr
L_0 g^{-1}(x)&=& z \partial_z g^{-1}(x) - \frac{1}{2} (\partial_\rho g^{-1}(x) - g^{-1}(x) (L_0 - \bar L_0) ) \cr
L_1 g^{-1}(x)&=& z^2 \partial_z g^{-1}(x) + z (- \partial_\rho g^{-1}(x) + g^{-1}(x) (L_0 - \bar L_0)) + e^{-\rho} g^{-1}(x)  (L_1 + \bar L_{-1})- e^{-2\rho} \partial_{\bar z} g^{-1}(x)  \cr
\bar L_{-1} g^{-1}(x)&=&  \partial_{\bar z} g^{-1}(x) \cr
\bar L_0 g^{-1}(x) &=& \bar z \partial_{\bar z} g^{-1}(x)  - \frac{1}{2} ( \partial_\rho g^{-1}(x) + g^{-1} (x) (L_0 -  \bar L_0)) \cr
\bar L_1 g^{-1}(x) &=& \bar z^2 \partial_{\bar z} g^{-1}(x)  -  {\bar z} (\partial_\rho g^{-1}(x)+ g^{-1}(x) (L_0 - \bar L_0)) + e^{-\rho} g^{-1}(x)  (L_{-1} + \bar L_1) - e^{-2\rho} \partial_{z} g^{-1}(x) \nonumber
\eea
Using these we can show as $\rho \rightarrow \infty$ that:
\bea
\langle h, \bar h; k, \bar k | L_{-1}  g^{-1}(x)| h, \bar h; j, -j \rangle\!\rangle &=& \partial_z \langle h, \bar h; k, \bar k   |g^{-1}(x) | h, \bar h; j, -j \rangle\!\rangle \cr
\langle h, \bar h; k, \bar k  |L_0 g^{-1}(x)  | h, \bar h; j, -j\rangle\!\rangle &=& (z \partial_z + h) \langle h, \bar h; k, \bar k   | g^{-1}(x) | h, \bar h; j, -j \rangle\!\rangle  \cr
\langle h, \bar h; k, \bar k  |L_1  g^{-1}(x) | h, \bar h; j, -j\rangle\!\rangle &=& (z^2 \partial_z + 2 z h) \langle h, \bar h; k, \bar k  | g^{-1}(x) | h, \bar h; j, -j \rangle\!\rangle  \cr
\langle h, \bar h; k, \bar k  |\bar L_{-1} g^{-1}(x)  | h, \bar h; j, -j \rangle\!\rangle &=&  \partial_{\bar z} \langle h, \bar h; k, \bar k   | g^{-1}(x) | h, \bar h; j, -j\rangle\!\rangle  \cr
\langle h, \bar h; k, \bar k  | \bar L_0 g^{-1}(x) | h, \bar h; j, -j\rangle\!\rangle &=&  (\bar z \partial_{\bar z} + \bar h)  \langle h, \bar h; k, \bar k   | g^{-1}(x) | h, \bar h; j, -j \rangle\!\rangle   \cr
\langle h, \bar h; k, \bar k  |\bar L_1 g^{-1}(x)  | h, \bar h; j, -j \rangle\!\rangle &=&   (\bar z^2 \partial_{\bar z} + 2 \bar z \bar h) \langle h, \bar h; k, \bar k   |g^{-1}(x) | h, \bar h; j, -j \rangle\!\rangle
\eea
This again means that the functions $\langle h, \bar h; k, \bar k   |g^{-1}(x) | h, \bar h; j, -j \rangle\!\rangle$ in the $\rho \rightarrow \infty$ limit also provide a representation of the algebra $sl(2, {\mathbb R}) \oplus sl(2, {\mathbb R})$. Again the (anti-) holomorphic part  has appeared in \cite{Bargmann:1946me, Jackiw:1990ka}. 

The last ingredient we want is the CG coefficients of unitary irreducible positive discrete series representations \cite{Bargmann:1946me} of $sl(2, {\mathbb R})$. These have been known for a long time \cite{holman} which we rework in the appendix \ref{appendixA} using our conventions. These are given as
\bea
C^{h_1h_2;h_3}_{k_1k_2;k_3} = \langle h_1, h_2; k_1, k_2 |h_1, h_2; h_3, k_3 \rangle = \frac{1}{\prod_{i=1}^3 \sqrt{k_i!(\Gamma(2h_i+k_i)}} f(k_1, k_2;k_3)
\eea
with $f(k_1, k_2; k_3)$ are as given by (\ref{ourcgs}) in the Appendix \ref{appendixA}.
%
%
We will not fix the normalisation as we do not need it in this paper. 

Finally we are ready to put together various components of our OWN diagrams with $N$ external legs with the corresponding representations $(h_i, \bar h_i)$ and compute them explicitly. The final answer will be proportion to $e^{-\rho ~ \sum_{i=1}^N h_i} \times e^{-\rho ~ \sum_{i=1}^N \bar h_i}$ times a function that is a product of a holomorphic part and an anti-holomorphic part. Let us now summarise the rules to compute the holomorphic part: 
\begin{itemize}
\item For each in-going external leg in representation $(h_i, \bar h_i)$ we associate the factor:
$$ \left[ i^{h_i} z_i^{-2h_i-k_i} \sqrt{\tfrac{\Gamma(2h_i+k_i)}{k_i! \Gamma(2h_i)}} \right] $$
\item For each out-going external leg in representation $(h_i, \bar h_i)$ associate the factor
$$\left[ (-i)^{h_i} z_i^{k_i} \sqrt{\tfrac{\Gamma(2h_i+k_i)}{k_i! \Gamma(2h_i)}} \right]  $$
\item For each trivalent vertex with two in-going (out-going) edges in representations $(h_m, \bar h_m)$,  $(h_n, \bar h_n)$ and one out-going (in-going) edge in the representation $(h_l, \bar h_l)$ we associate a CG coefficient $C^{h_m h_n; h_l}_{k_m k_n; k_l}$.
\item Finally sum over all repeated $k_i$s.
\end{itemize}
The rules to compute the anti-holomorphic factor in the OWN are simply obtained from the above ones by replacing $h_i \rightarrow \bar h_i$, $k_i \rightarrow \bar k_i$ and then complex conjugating the rest. The boundary CFT answers are the same as the OWN answers but for the $\rho$-dependent pre-factors.

\subsection{The 2-point function recovered}
The 2-point function of two primary operators in a CFT is completely determined by the symmetries and therefore it is our simplest partial wave. We should be able to derive it from the simplest OWN which is just a line with the end points approaching the boundary. 
\begin{center}
\setlength{\unitlength}{1.25cm}
\begin{picture}(3,3)(0,-1.25)
\put(1,1){\vector(2,0){1.25}}
\put(1,1){\line(2,0){2.5}}
\put(.8,.7){$x_1$}
\put(3.5,.7){$x_2$}
\put(1.45,1.1){\scalebox{0.7}{$(h, \bar h)$}}
\put(-.5,1){\scalebox{0.7}{$\langle\!\langle h \bar h; j, j |$}}
\put(4,1){\scalebox{0.7}{$|h, \bar h; j, -j \rangle\!\rangle$}}
\put(-.6,-.5){{\bf Fig. \!\!2}: Spin network for 2-point function}
\end{picture}
\end{center}
According to our prescription it is given by
\bea
&& \langle\!\langle h, \bar h; j, j | g(x_1) g^{-1}(x_2)|h, \bar h; j,-j \rangle\!\rangle \cr
&=& \sum_{k, \bar k = 0}^\infty \langle\!\langle h, \bar h; j, j | g(x_1) |h, \bar h; k, \bar k \rangle \langle h, \bar h; k, \bar k| g^{-1}(x_2)|h, \bar h; j,-j \rangle\!\rangle \cr
&{\rho_1 = \rho_2 = \rho \rightarrow \infty \atop =}& \lambda^2 \frac{(-1)^{-j} \Gamma(2\bar h)}{\Gamma(2h) (2j)!} z_1^{-2h} \bar z_1^{-2\bar h} \sum_{k, \bar k =0}^\infty \frac{\Gamma(2h+k)}{k! \Gamma(2h)} \left(\frac{z_2}{z_1} \right)^k \frac{\Gamma(2 \bar h+ \bar k)}{\bar k! \Gamma(2\bar h)}\left(\frac{\bar z_2}{ \bar z_1} \right)^{\bar k} \cr
& =& \frac{e^{-2\rho (	h+ \bar h)}}{(z_1-z_2)^{2h} (\bar z_1 - \bar z_2)^{2\bar h}}
\eea
where we have used the value of $\lambda$ as in (\ref{lambda}). Therefore the correctly normalised 2-point function is obtained by taking 
\bea
\langle {\cal O}_{(h, \bar h)} (z_1, \bar z_1) {\cal O}_{(h, \bar h)}(z_2, \bar z_2) \rangle &=& e^{2\rho(h+\bar h)}  \langle\!\langle h, \bar h; j, j| g(z_1, \bar z_1, \rho) g^{-1}(z_2, \bar z_2, \rho)|h, \bar h; j,-j \rangle\!\rangle \big{|}_{\rho \rightarrow \infty} \cr
&=& \frac{1}{(z_1-z_2)^{2h} (\bar z_1 - \bar z_2)^{2\bar h}}
\eea
It should be clear that when we consider the 2-point function of primaries with different conformal dimensions the corresponding Wilson line vanishes as the Wilson line operator does not change the representation of the state $\langle\!\langle h \bar h; j, j |$ ($|h, \bar h; j, -j \rangle\!\rangle$) when it acts to the left (right) and the resultant overlap simply vanishes as the representations $(h_1, \bar h_1)$ and $(h_2, \bar h_2)$ will be orthogonal when $h_1 \ne h_2$ or $\bar h_1 \ne \bar h_2$.

Note that this computation suggests that the dual to the state $\langle\!\langle h, \bar h; j, m |$ should be taken to be $|h, \bar h; j, -m \rangle\!\rangle$. This is not an unreasonable choice as the conformal transformation that takes the representation provided by the functions  $\langle h, \bar h; k, \bar k| g^{-1}(x) |h, \bar h; j, m \rangle \!\rangle $ to the representation provided by $\langle h, \bar h; k, \bar k| g^{-1}(x) |h, \bar h; j, m \rangle \!\rangle $ is $z \rightarrow -1/z$, $\bar z \rightarrow -1/\bar z$. In polar coordinates on the complex plane this is $r \rightarrow 1/r$ which corresponds to the time-reversal operation on the cylinder under the state-operator correspondence. It is well known that the time-reversal operation acts on angular momentum eigenstates in this fashion.

\subsection{The 3-point function recovered}
We can now turn to computing the 3-point function which is also a partial wave on its own. For this we consider a three pronged Open Wilson Network as given below:
\begin{center}
\setlength{\unitlength}{1.25cm}
\begin{picture}(5,5)(-1.5,-2)
\put(1,1){\vector(-1,-1){.5}}
\put(0,0){\line(1,1){.5}}
\put(0,2){\line(1,-1){1}}
\put(1,1){\vector(-1,1){.5}}
\put(2.5,1){\vector(-2,0){.75}}
\put(1,1){\line(2,0){1.5}}
\put(.42,1.8){\begin{turn}{-45}{\scalebox{0.7}{$(h_1, \bar h_1)$}}\end{turn}}
\put(.35,.15){\begin{turn}{45}{\scalebox{0.7}{$(h_2, \bar h_2)$}}\end{turn}}
\put(1.5,1.1){\scalebox{0.7}{$(h_3, \bar h_3)$}}
\put(-1.55,-.15){\scalebox{0.7}{$\langle\!\langle h_2, \bar h_2; j_2, -j_2 |$}}
\put(-1.5,2){\scalebox{0.7}{$\langle\!\langle h_1 \bar h_1; j_1, -j_1 |$}}
\put(2.5,.95){\scalebox{0.7}{$|h_3, \bar h_3; j_3, j_3 \rangle\!\rangle$}}
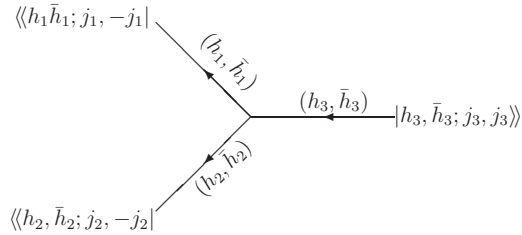
\put(-2.2,-1.5){{\bf Fig. \!\!3}: Spin network for CFT 3-point function}
\end{picture}
\end{center}
Following our prescription we associate the following answer to this diagram:
\bea
&& \sum_{k_i = 0}^\infty \left[ (-1)^{-\frac{h_1}{2}} e^{-\rho h_1} z_1^{k_1} \sqrt{\frac{\Gamma(2h_1+k_1)}{k_1! \Gamma(2h_1)}} \right]   \times \left[(-1)^{-\frac{h_2}{2}} e^{-\rho h_2} z_2^{k_2} \sqrt{\frac{\Gamma(2h_2+k_2)}{k_2! \Gamma(2h_2)}} \right] \cr
&& \times  \left[ (-1)^{\frac{h_3}{2}} e^{-\rho h_3} z_3^{-2h_3-k_3} \sqrt{\frac{\Gamma(2h_3+k_3)}{k_3! \Gamma(2h_3)}} \right]\times C^{h_1h_2h_3}_{k_1k_2k_3} \cr
&& \times \sum_{\bar k_i = 0}^\infty \left[ (-1)^{\frac{\bar h_1}{2}} e^{-\rho \bar h_1} \bar z_1^{\bar k_1} \sqrt{\frac{\Gamma(2\bar h_1+ \bar k_1)}{\bar k_1! \Gamma(2 \bar h_1)}} \right]   \times \left[(-1)^{\frac{\bar h_2}{2}} e^{-\rho \bar h_2} \bar z_2^{\bar k_2} \sqrt{\frac{\Gamma(2 \bar h_2+  \bar k_2)}{ \bar k_2! \Gamma(2 \bar h_2)}} \right] \cr
&& \times  \left[ (-1)^{-\frac{\bar h_3}{2}} e^{-\rho \bar h_3} \bar z_3^{-2\bar h_3-\bar k_3} \sqrt{\frac{\Gamma(2\bar h_3+\bar k_3)}{\bar k_3! \Gamma(2\bar h_3)}} \right]\times C^{\bar h_1\bar h_2\bar h_3}_{\bar k_1\bar k_2\bar k_3}
\eea
Clearly the answer is a product of holomorphic and anti-holomorphic pieces each of which can be computed separately. The holomorphic part can be summed to obtain:
\bea
&& \sum_{k_i = 0}^\infty \left[ (-1)^{-\frac{h_1}{2}} e^{-\rho h_1} z_1^{k_1} \sqrt{\frac{\Gamma(2h_1+k_1)}{k_1! \Gamma(2h_1)}} \right]   \times \left[(-1)^{-\frac{h_2}{2}} e^{-\rho h_2} z_2^{k_2} \sqrt{\frac{\Gamma(2h_2+k_2)}{k_2! \Gamma(2h_2)}} \right] \cr
&& \times  \left[ (-1)^{\frac{h_3}{2}} e^{-\rho h_3} z_3^{-2h_3-k_3} \sqrt{\frac{\Gamma(2h_3+k_3)}{k_3! \Gamma(2h_3)}} \right]\times C^{h_1h_2h_3}_{k_1k_2k_3}\cr
&\sim & \frac{1}{(z_1-z_2)^{h_1+h_2 - h_3} (z_2 - z_3)^{h_2+h_3 - h_1} (z_3 - z_1)^{h_3+h_1 - h_2}}
\eea
The details of this calculation are relegated to the appendix \ref{appendixc}. Similarly the the anti-holomorphic part will give an answer proportional to 
\bea
\frac{1}{(\bar z_1- \bar z_2)^{\bar h_1+ \bar h_2 - \bar h_3} ( \bar z_2 - \bar  z_3)^{\bar h_2+\bar h_3 - \bar h_1} (\bar z_3 -\bar z_1)^{\bar h_3+\bar h_1 - \bar h_2}}
\eea
Multiplying both these factors together one recovers the precise coordinate behaviour of the 3-point function of primaries in the CFT.
\subsection{The 4-point partial wave recovered}
For this we consider the OWN in Fig.(4) below:
\begin{center}
\setlength{\unitlength}{1.25cm}
\begin{picture}(5,5)(-0.5,-2)
\put(0,0){\vector(1,1){.5}}
\put(.5,.5){\line(1,1){.5}}
\put(1,1){\line(-1,1){1}}
\put(0,2){\vector(1,-1){.5}}
\put(1,1){\vector(2,0){1}}
\put(1,1){\line(2,0){2}}
\put(3,1){\line(1,1){1}}
\put(3,1){\vector(1,1){.5}}
\put(3,1){\line(1,-1){1}}
\put(3,1){\vector(1,-1){.5}}
\put(.35,1.75){\begin{turn}{-45}{\scalebox{0.7}{$(h_1, \bar h_1)$}}\end{turn}}
\put(.35,.15){\begin{turn}{45}{\scalebox{0.7}{$(h_2, \bar h_2)$}}\end{turn}}
\put(3,.5){\begin{turn}{-45}{\scalebox{0.7}{$(h_3, \bar h_3)$}}\end{turn}}
\put(3,1.3){\begin{turn}{45}{\scalebox{0.7}{$(h_4, \bar h_4)$}}\end{turn}}
\put(1.6,1.1){\scalebox{0.7}{$(h, \bar h)$}}
\put(-1.5,2){\scalebox{0.7}{$\langle\!\langle h_1, \bar h_1; j_1; j_1  |$}}
\put(-1.5,-.2){\scalebox{0.7}{$\langle\!\langle h_2, \bar h_2; j_2; j_2 |$}}
\put(4,-.2){\scalebox{0.7}{$|h_3, \bar h_3; j_3; -j_3 \rangle\!\rangle$}}
\put(4,2){\scalebox{0.7}{$|h_4, \bar h_4; j_4; -j_4\rangle\!\rangle$}}
\put(-.5,-1.5){{\bf Fig. \!\!4}: Partial Wave of 4-point function}
\label{Fig4}
\end{picture}
\end{center}
whose answer is 
\bea
\label{eq:spinning_block}
\sum_{k_i,\bar k_i} && \langle\!\langle h_1, \bar h_1; j_1,j_1|g(x_1)|h_1, \bar h_1; k_1,\bar k_1 \rangle 
 \times \langle\!\langle h_2, \bar h_2; j_2, j_2|g(x_2)|h_2, \bar h_2; k_2,\bar k_2 \rangle\cr
 && \times \langle h_3, \bar h_3; k_3,\bar k_3|g^{-1}(x_3)|h_3, \bar h_3; j_3,-j_3 \rangle\!\rangle \times
\langle h_4, \bar h_4; k_4,\bar k_4|g^{-1}(x_4)|h_4, \bar h_4; j_4,-j_4 \rangle\!\rangle \cr
&& \cr
&& \times \sum_{k,\bar k} C^{h_1h_2h}_{k_1k_2k} \times C^{h h_3h_4}_{kk_3k_4} \times C^{\bar h_1\bar h_2\bar h}_{\bar k_1\bar k_2\bar k} \times C^{\bar h \bar h_3 \bar h_4}_{\bar k \bar k_3 \bar k_4} 
\eea
where $x_i = (z_i, \bar z_i, \rho \rightarrow \infty)$. One can in principle compute this quantity and recover the full coordinate dependence of this 4-point partial wave (as guaranteed by the differential relations we had established earlier). However to simplify the presentation and as it is standard we take $z_1 \rightarrow \infty$, $z_2 \rightarrow 1$, $z_3 \rightarrow z$ and $z_4 \rightarrow 0$. Then the partial wave in the decomposition of the 4-point function of four primaries of dimensions $(h_i, \bar h_i)$ for $i=1,2,3,4$ takes the form: 
\bea
\label{cftW4}
W_{(h,\bar h)}(x) = z_1^{-2h_1}\bar z_1^{- 2\bar h_1}  G_{(h, \bar h)}(z)
\eea
%
%
From the CFT it is known that $G$ takes the following form:
\bea
\label{cftG4}
G_{(h, \bar h)}(z, \bar z) \sim z^{-h_3-h_4} \bar z^{-\bar h_3-\bar h_4}  {\cal F}_h(z) \bar {\cal F}_{\bar h}(\bar z)
\eea
where ${\cal F}$ and $\bar {\cal F}$ are supposed to be the corresponding conformal blocks. We can now compute using our prescription this partial wave and hence the blocks.
%
%
 When we set $z_1 \rightarrow \infty$, $z_2 \rightarrow 1$, $z_3 \rightarrow z$ and $z_4 \rightarrow 0$ then each component in the above expression simplifies as follows:
\bea
 \langle\!\langle h_1, \bar h_1; j_1,j_1|g(x_1)|h_1, \bar h_1; k_1,\bar k_1 \rangle &\rightarrow& (-1)^{\frac{h_1}{2}} e^{-\rho h_1} z_1^{-2h_1} \delta_{k_1,0} \times (-1)^{-\frac{\bar h_1}{2}} e^{-\rho \bar h_1} \bar z_1^{-2 \bar h_1} \delta_{\bar k_1,0} \cr
  \langle\!\langle h_2, \bar h_2; j_2, j_2|g(x_2)|h_2, \bar h_2; k_2,\bar k_2 \rangle &\rightarrow& (-1)^{\frac{h_2}{2}} e^{-\rho h_2} \sqrt{\tfrac{\Gamma(2h_2 + k_2)}{k_2! \Gamma(2h_2)}} \times (-1)^{-\frac{\bar h_2}{2}} e^{-\rho \bar h_2} \sqrt{\tfrac{\Gamma(2\bar h_2 + \bar k_2)}{\bar k_2! \Gamma(2 \bar h_2)}} \cr
  \langle h_3, \bar h_3; k_3,\bar k_3|g^{-1}(x_3)|h_3, \bar h_3; j_3, -j_3 \rangle\!\rangle &\rightarrow& (-1)^{-\frac{h_3}{2}} e^{-\rho h_3} z^{k_3} \sqrt{\tfrac{\Gamma(2h_3+k_3)}{k_3! \Gamma(2h_3)}}  \times (-1)^{\frac{\bar h_3}{2}} e^{-\rho \bar h_3} \bar z^{\bar k_3} \sqrt{\tfrac{\Gamma(2\bar h_3+\bar k_3)}{\bar k_3! \Gamma(2\bar h_3)}} \cr
  \langle h_4, \bar h_4; k_4,\bar k_4|g^{-1}(x_4)|h_4, \bar h_4; j_4, -j_4 \rangle\!\rangle &\rightarrow& (-1)^{-\frac{h_4}{2}} e^{-\rho h_4} \delta_{k_4,0} \times (-1)^{\frac{\bar h_4}{2}} e^{-\rho \bar h_4} \delta_{\bar k_4,0}
\eea
%
%
%
We now need the CG coefficients which again fall into different cases. Let us first assume that we have $h_1+h_2 \ge h$ and $h_3 + h_4 \ge h$. In this case the relevant CG coefficients are
\begin{align}
 C^{h_1,h_2,h}_{0,k_2,k} &\sim \delta_{h_1+h_2+k_2-h-k}\times \tfrac{\Gamma(k-k_2)}{\sqrt{\Gamma(2 h_1)}}\times \sqrt{\tfrac{k_2!\Gamma(2h_2+k_2)}{k!\Gamma(2h+k)}} \\
 C^{h,h_3,h_4}_{k,k_3,0} &\sim \delta_{h_3+h_4+k_3-h-k}\times \tfrac{\Gamma(k-k_3)}{\sqrt{\Gamma(2 h_4)}}\times \sqrt{\tfrac{k_3!\Gamma(2h_3+k_3)}{k!\Gamma(2h+k)}} 
\end{align}
The holomorphic part of (\ref{eq:spinning_block}) becomes
\bea
& \sim & (-1)^{\frac{1}{2}(h_1+h_2- h_3-h_4)} e^{-\rho (h_1+h_2 + h_3 + h_4)} z^{-2h_1} \sum_{k_i, k} \sqrt{\tfrac{\Gamma(2h_2+k_2)}{k_2! \Gamma(2h_2)}} \sqrt{\tfrac{\Gamma(2h_3+k_3)}{k_3! \Gamma(2h_3)}} z^{k_3} \cr
&& \times \delta_{h_1 + h_2 - h + k_2 - k}  \tfrac{\Gamma(k-k_2)}{\sqrt{\Gamma(2 h_1)}}\times \sqrt{\tfrac{k_2!\Gamma(2h_2+k_2)}{k!\Gamma(2h+k)}}  
\times \delta_{h_3+h_4+k_3-h-k}\times \tfrac{\Gamma(k-k_3)}{\sqrt{\Gamma(2 h_4)}}\times \sqrt{\tfrac{k_3!\Gamma(2h_3+k_3)}{k!\Gamma(2h+k)}} \cr
&\sim & e^{-\rho (h_1+h_2 + h_3 + h_4)} z_1^{-2h_1} \sum_{k = 0}^\infty \tfrac{\Gamma(h-h_1+h_2+k) \Gamma(h+h_3-h_4+k)}{\Gamma(2h+k)} z^{k} \cr
& \sim&  e^{-\rho (h_1+h_2 + h_3 + h_4)} z_1^{-2h_1} z^{h-h_3-h_4} {}_2F_1[h-h_1+h_2, h+h_3-h_4, 2h, z] \cr &&
\eea 
where in the second step we carried out the sums over $k_2$ and $k_3$ using the Kronecker deltas. One gets for the anti-holomorphic part 
\bea
\sim  e^{-\rho (\bar h_1+ \bar h_2 + \bar h_3 + \bar h_4)} \bar z_1^{-2\bar h_1} \bar z^{\bar h-\bar h_3-\bar h_4} {}_2F_1[\bar h-\bar h_1+ \bar h_2, \bar h+ \bar h_3- \bar h_4, 2 \bar h, \bar z].
\eea
Comparing our answer with (\ref{cftW4}, \ref{cftG4}) we recover the well-known answer \cite{Perlmutter:2015iya} for the 4-point spinning global conformal block
\bea
\label{4ptblocks}
{\cal F}_h(z) &\sim& z^h ~ {}_2F_1[h-h_1+h_2, h+h_3-h_4, 2h, z]\cr
 \bar {\cal F}_{\bar h}(\bar z) &\sim& \bar z^{\bar h} ~ {}_2F_1[\bar h-\bar h_1+ \bar h_2, \bar h+ \bar h_3- \bar h_4, 2 \bar h, \bar z]
\eea
As mentioned above in this calculation we have assumed $h\le h_3+h_4$ and $h\le h_1+h_2$. There are three other possibilities which can also be computed easily using the appropriate expressions of the CG coefficients (\ref{ourcgs}) to give answers exactly of the same form.

Note that what we have computed satisfies two independent conformal Casimir equations -- one for each of the two $sl(2, {\mathbb R})$ factors in the 2d global conformal algebra with eigenvalues $2h(h-1)$ and $2\bar h (\bar h-1)$ respectively. The global partial wave however is supposed to satisfy one conformal Casimir equation with the Casimir operator given by the sum of these two Casimirs with eigenvalue $2h(h-1) + 2\bar h (\bar h-1)$. This eigenvalue is invariant under $h \leftrightarrow \bar h$. The OWN considered above continues to be a solution to this one Casimir equation. But there is a second independent solution with the same eigenvalue obtained from the above OWN by interchanging $h$ with $\bar h$. Therefore any linear combination of these two OWNs would provide a solution to the conformal Casimir equation. A basis in this space of solutions can be taken to be the symmetric and the antisymmetric combinations under $h \leftrightarrow \bar h$. As advocated, say, in \cite{Dolan:2011dv} the symmetric combination is the one satisfying the appropriate boundary conditions. This in our context gives us the $G_{\Delta, l} (z, \bar z)$ given by $z^{-h_3-h_4} \bar z^{-\bar h_3- \bar h_4}$ times :
\bea
 |z|^{\Delta - l} \left[z^l ~ {}_2F_1[\tfrac{\Delta-l}{2} - h_{12}, \tfrac{\Delta-l}{2} - h_{34}, \Delta-l, z] {}_2F_1[\tfrac{\Delta+l}{2} - h_{12}, \tfrac{\Delta+l}{2} - h_{34}, \Delta+l, \bar z] + (z \rightarrow \bar z, h_{ij} \rightarrow \bar h_{ij}) \right] \cr &&
\eea
where $h_{ij} = h_i - h_j$ etc. and $\Delta = h+\bar h$, $l = h - \bar h$. This is our final answer for the 4-point partial wave of primaries and clearly matches with (\ref{4ptscalarblock}) when we take $h_i  = \bar h_i$ as it is appropriate for scalar operators in the external legs and satisfies the same boundary conditions as $z, \bar z \rightarrow 0$.\footnote{We have considered in here the case of the boundary conditions imposed when cross ratios $(z, \bar z)$ approach zero. One can similarly consider diagrams that compute blocks with boundary conditions imposed as $(z, \bar z)$ approaches $(1,1)$ or $(\infty, \infty)$.}

\subsection{The 5-point conformal block recovered}
The last example we consider here is the conformal partial wave that appears in the pants decomposition of the 5-point function of primaries. For this we consider the following Open Wilson Network (Fig. 5).
%
\begin{center}
\setlength{\unitlength}{1.4cm}
\begin{picture}(5,5)(0,-1.5)
\put(0,0){\vector(1,0){.75}}
\put(0,0){\line(1,0){1.5}}
\put(-.15,-.25){$\infty$}
\put(1.5,1.5){\line(0,-1){1.5}}
\put(1.5,1.5){\vector(0,-1){.75}}
\put(1.3,1,4){$1$}
\put(1.5,0){\line(1,0){1.5}}
\put(1.5,0){\vector(1,0){.75}}
\put(3,1.5){\line(0,-1){1.5}}
\put(3,1.5){\vector(0,-1){.75}}
\put(2.75,1.4){$z_3$}
\put(3,0){\line(1,0){1.5}}
\put(3,0){\vector(1,0){.75}}
\put(4.5,0){\vector(1,0){.75}}
\put(4.5,0){\line(1,0){1.5}}
\put(4.5,1.5){\line(0,-1){1.5}}
\put(4.5,0){\vector(0,1){.75}}
\put(4.25,1.4){$z_4$}
\put(6,-.25){$0$}
\put(-1.75,-.1){\scalebox{0.7}{$\langle\!\langle h_1, \bar h_1; j_1; j_1 |$}}
\put(.35,.15){\scalebox{0.7}{$(h_1, \bar h_1)$}}
\put(1.6,.5){\begin{turn}{90}{\scalebox{0.7}{$(h_2, \bar h_2)$}}\end{turn}}
\put(3.1,.5){\begin{turn}{90}{\scalebox{0.7}{$(h_3, \bar h_3)$}}\end{turn}}
\put(1.85,.15){\scalebox{0.7}{$(h, \bar h)$}}
\put(3.35,.15){\scalebox{0.7}{$(h', \bar h')$}}
\put(4.6,.5){\begin{turn}{90}{\scalebox{0.7}{$(h_4, \bar h_4)$}}\end{turn}}
\put(4.85,.15){\scalebox{0.7}{$(h_5, \bar h_5)$}}
\put(6.25,-.1){\scalebox{0.7}{$|h_5, \bar h_5; j_5; -j_5 \rangle\!\rangle$}}
\put(1.35,1.75){\begin{turn}{90}{\scalebox{0.7}{$|h_2, \bar h_2; j_2; j_2\rangle\!\rangle$}}\end{turn}}
\put(2.85,1.75){\begin{turn}{90}{\scalebox{0.7}{$|h_3, \bar h_3; j_3; j_3\rangle\!\rangle$}}\end{turn}}
\put(4.35,1.75){\begin{turn}{90}{\scalebox{0.7}{$|h_3, \bar h_4; j_4; -j_4\rangle\!\rangle$}}\end{turn}}
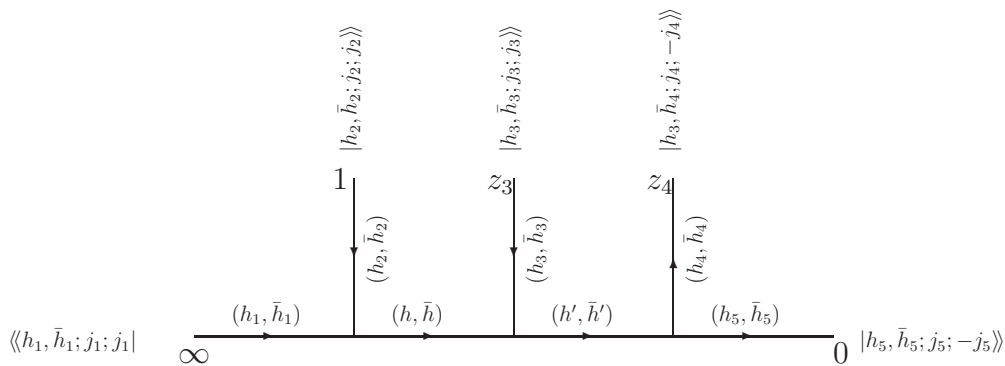
\put(-.5,-1.5){{\bf Fig. \!\!5}: A Partial Wave of 5-point function}
\end{picture}
\end{center}

%
The value of the holomorphic part of this diagram is up to a factor $e^{-\rho (h_1+h_2+h_3+h_4+h_5)}$:
\bea
&\sim & z_3^{-2h_3} \sum_{k_2, k_3, k_4, k, k'} C^{h_1h_2;h}_{0 k_2; k} \sqrt{\tfrac{\Gamma(2h_2+k_2)}{k_2! \Gamma(2h_2)}} C^{hh_3;h'}_{kk_3;k'} \sqrt{\tfrac{\Gamma(2h_3+k_3)}{k_3! \Gamma(2h_3)}}  C^{h_4h_5;h'}_{k_4 0; k'} \sqrt{\tfrac{\Gamma(2h_4+k_4)}{k_4! \Gamma(2h_4)}} ~ z_3^{-k_3} z_4^{k_4}
\eea
Let us further assume that $h_1+h_2 \ge h$, $h+h_3 \ge h'$ and $h_4+h_5 \ge h'$.\footnote{All the other possibilities can also be considered and computed with the suitable CG coefficients resulting in expressions with the same coordinate dependence.} The CG coefficients are given by
\bea
 C^{h_1,h_2,h}_{0,k_2,k} &\sim \delta_{h_1+h_2+k_2-h-k} ~ \tfrac{\Gamma(k-k_2)}{\sqrt{\Gamma(2 h_1)}} ~ \sqrt{\tfrac{k_2!\Gamma(2h_2+k_2)}{k!\Gamma(2h+k)}} \cr
 C^{h_4h_5;h'}_{k_4 0; k'} &\sim \delta_{h_4+h_5+k_4-h'-k'}~ \tfrac{\Gamma(k'-k_4)}{\sqrt{\Gamma(2 h_5)}}~ \sqrt{\tfrac{k_4!\Gamma(2h_4+k_4)}{k'!\Gamma(2h'+k')}}
\eea
and
\bea
 C^{hh_3;h'}_{kk_3;k'} =  \delta_{h+h_3-h'+k+k_3-k'}\tfrac{\Gamma(k'-k_3)k_3!\Gamma(2h_3+k+k_3)\Gamma(h+h'-h_3)}{\sqrt{k!k_3!k'!\Gamma(2h+k)\Gamma(2h_3+k_3)\Gamma(2h'+k')}}\cr
                         \times {}_3F_2 \left({{-k,-k',2h+k+k_3-k'}\atop{1+k_3-k',1-2h_3-k-k_3}};1\right)
\eea
Then the value of 5-point block thus becomes
\bea
 &&z_3^{-2h_3}\sum_{k,k'} \sum_{k_2,k_3,k_4} z_3^{-k_3} z_4^{k_4} ~~  {}_3F_2 \left({{-k,-k',2h+k+k_3-k'}\atop{1+k_3-k',1-2h_3-k-k_3}};1\right) \cr
 \cr
 && \times \tfrac{\Gamma(k-k_2)\Gamma(k'-k_4)\Gamma(k'-k_3)\Gamma(2h_3+k+k_3)\Gamma(2h_2+k_2)\Gamma(2h_4+k_4)\Gamma(h+h'-h_3)}{k!k'!\Gamma(2h+k)\Gamma(2h'+k')} \cr
 && \times \delta_{h_1+h_2+k_2-h-k}~ \delta_{h_4+h_5+k_4-h'-k'} ~ \delta_{h+h_3-h'+k+k_3-k'}
\eea
The 5-point global block has been computed recently using the CFT methods in \cite{Alkalaev:2015fbw} and to compare with their answer we write above expression as
\bea
 &&z_3^{-2h_3}z_3^{h-h_3-h'}z_4^{h'-h_4-h_5}\sum_{k,k'} \sum_{k_2,k_3,k_4}q_1^{-h_3-k_3+h_4+k_4+h_5-h} q_2^{h_4+k_4+h_5-h'} \cr
 &&\times  {}_3F_2 \left({{-k,-k',2h+k+k_3-k'}\atop{1+k_3-k',1-2h_3-k-k_3}};1\right) \cr
 \cr
 && \times \tfrac{\Gamma(k-k_2)\Gamma(k'-k_4)\Gamma(k'-k_3)\Gamma(2h_3+k+k_3)\Gamma(2h_2+k_2)\Gamma(2h_4+k_4)\Gamma(h+h'-h_3)}{k!k'!\Gamma(2h+k)\Gamma(2h'+k')} \cr
 && \times \delta_{h_1+h_2+k_2-h-k} ~ \delta_{h+h_3-h'+k+k_3-k'} ~ \delta_{h_4+h_5+k_4-h'-k'}
\eea
where $q_1=z_3$ and $q_2=z_4/z_3$.
Then we do $k_2,k_3,k_4$ sums using 1st, 2nd and 3rd Kronecker deltas respectively in the above expression. The result becomes
\bea
 &&\sim z_3^{-2h_3}z_3^{h-h_3-h'}z_4^{h'-h_4-h_5} \sum_{k, k'} \frac{q_1^{k} {q_2}^{k'}}{k! k'!} \cr
   \cr
   && \times \tfrac{\Gamma(-h_1+h+h_2+k)\Gamma(h-h'+h_3+k)\Gamma(-h+h'+h_3+k')\Gamma(h'+h_4-h_5+k')}{ \Gamma (2 h+k) \Gamma (2h'+k')} \cr
   \cr
   &&\times  _3F_2 \left({{-k,-k',h+h'-h_3} \atop{h-h'-h_3-k'+1,-h+h'-h_3-k+1}};1\right)
\eea
The hypergeometric function here can be rewritten using the (Shepperd's) identity \cite{aar1999}
\bea
{}_3F_2 \left({{-n,a,b}\atop{d,e}};1 \right) = \frac{(e-a)_n}{(e)_n}{}_3F_2 \left({{-n,a,d-b}\atop{d,a+1-n-e}};1 \right)
\eea
as
\bea
&&{}_3F_2 \left({{-k,-k',h+h'-h_3} \atop{h-h'-h_3-k'+1,-h+h'-h_3-k+1}};1\right) \cr
\cr
&& = \frac{\Gamma(h-h'+h_3+k-k')\Gamma(h-h'+h_3)}{\Gamma(h-h'+h_3-k')\Gamma(h-h'+h_3+k)} 
~~ {}_3F_2 \left({{-k,-k',-2h'-k'+1} \atop{h-h'-h_3-k'+1,h_3+h-h'-k'}};1\right) \cr &&
\eea
Finally we get the holomorphic part of the 5-point block in the form
\bea
 \sim z_3^{-2h_3}z_3^{h-h_3-h'}z_4^{h'-h_4-h_5} \sum_{k, k'=0}^{\infty} F_{k,k'}q_1^{k} {q_2}^{k'}
\eea
where
\bea
 F_{k,k'} = \frac{1}{k!k'!} \frac{\Gamma(h+h_2-h_1+k)\Gamma(h'+h_4-h_5+k')}{\Gamma(2h+k)\Gamma(2h'+k')}\tau_{k,k'}
\eea
with
\bea
 \tau_{k,k'} &&= \frac{\Gamma(h_3-h+h'+k')\Gamma(h_3+h-h'+k-k')}{\Gamma(h_3+h-h'-k')} \cr
 \cr
 && \times {}_3F_2 \left({{-k,-k',-2h'-k'+1} \atop{h-h'-h_3-k'+1,h_3+h-h'-k'}};1\right)
\eea
which apart from a purely $h_i$-dependent pre-factor is exactly identical to the one obtained in \cite{Alkalaev:2015fbw}. The anti-holomorphic part can also be computed on similar lines and put together with the holomorphic part to find the contribution of the OWN in Fig.(5) to the 5-point partial wave.

The 5-point block is a solution to two pairs of Casimir equations (two each for each of the two intermediate edges) as discussed earlier. However just as in the case of the 4-point partial wave we need impose only two Casimir equations. Then we have four OWN diagrams related to the one in Fig.(5) under $(h \leftrightarrow \bar h)$ or $(h' \leftrightarrow \bar h')$ all of which solve these two equations. Again the generic solution would be a linear combination of all four solutions and one would have to pick appropriate combinations depending on the boundary conditions one imposes. 

Now that we have demonstrated our method at work successfully one can in principle compute straightforwardly the higher point (global) blocks as well as the partial waves for a given decomposition of that higher point function of primaries.

\section{Discussion}

We have considered open Wilson network operators that can be defined using the gauge connection (\ref{gaugeconnection}) made of the spin-connections and the vielbeins in the first order formulation of gravity with negative cosmological constant and showed that with their end points on the boundary they compute the (global) partial waves and conformal blocks of the dual CFT. 

We restricted to computing the expectation values of the OWN operators in the semiclassical limit by simply evaluating them in the background of the AdS space. We have shown that they do satisfy the expected differential equations and developed methods to compute them. The computations in $d=2$ case are presented explicitly till the 5-point block involving primary operators of arbitrary spin and shown to agree exactly with the previously known results. Our results should be seen as complementary to the ones of \cite{Hijano:2015zsa}.

Note that our prescription to compute conformal blocks and partial waves did not require putting any input of the actual interactions of the relevant bulk fields. One should have expected this as these quantities are determined solely by the symmetries and not the dynamics of the CFT. It should be pointed out that at the end our computations in the beginning of section 3 resulted in a prescription to obtain the global blocks which didn't really involve the bulk. Various components involved in the construction could have been obtained simply on the boundary CFT side. This must also be seen as a reflection of the kinematical nature of the conformal blocks.

People have used particle description to compute (at least in the classical limit, see for instance, \cite{Ammon:2013hba, Hijano:2015rla}) Wilson line operator expectation values in the context of computing the entanglement entropy and classical blocks. It will be interesting to find the translation between the particle language and ours in terms of open Wilson networks.

We have not normalised the CG coefficients as we have been interested in computing the dependence on the cross ratios first. One expects additional phases to appear when one goes from one orientation to another for a given graph as well as under permutations of the end points. This may be relevant when one considers spin networks with braiding and knots \cite{Witten:1989wf}.

We have restricted to computing only the global blocks. One should like to be able to compute the Virasoro blocks and in different backgrounds relevant to heavy-light and classical limits of the Virasoro blocks  \cite{Hijano:2015qja}. The quantum corrections to the classical answers provided here should correspond to computing sub-leading corrections to the large-$c$ limit \cite{Hijano:2015rla}. Graphs with the same number of external legs but more complicated internal structure (such as loops etc.) may conceivably be relevant in the computation of the quantum corrections. In this connection the relation to $sl(2,{\mathbb C})_q$ will be interesting to explore \cite{Verlinde:2015qfa} (see also \cite{Chun:2015gda}).

We can in principle consider blocks with external operators being descendants of primaries. We expect that the OWNs with $m$ values other than $m=j$ (for ingoing) or $m=-j$ (for outgoing) to correspond to some of these. The question of how to compute partial waves for the Virasoro descendants of primaries remains open. It is possible that one has to consider cap states to belong to representations of the twisted diagonal combination of the two Virasoro algebras that emerge as the asymptotic symmetry algebra of the bulk theory as suggest in \cite{Verlinde:2015qfa} in the context of bulk local fields. It will be interesting to explore this issue further. 

We took the boundary to be ${\mathbb R}^2$ in section 3. We can easily take it to be either $S^2$, ${\mathbb H}^2$ or $S^1 \times {\mathbb R}$. Also one could consider expectation values of the OWNs in a highly excited state by replacing the bulk with the appropriate geometry (such as a conical singularity or a BTZ black hole). It will be interesting to consider the case of the boundary being any Riemann surface as well.

Most of what we have presented here can be generalised to dimensions higher than three (i.e, $d>2$) and we hope to present the results elsewhere soon \cite{BRS2}. Our original motivation is the case of higher spin theories in three and higher dimensions. We would like to generalise our construction to address questions in these theories on our lines. The authors of \cite{deBoer:2013vca, deBoer:2014sna} have considered wilson lines in these theories in the context of holographic entanglement entropy and the conformal blocks. We hope to be able to report progress on this front in the near future.

\section*{Acknowledgements}
We would like to thank Ghanashyam Date, Romesh Kaul, Gautam Mandal, Rohan Poojary and Ashoke Sen for helpful conversations and interest.
\appendix
\section{Some Group Theory}
\label{appendixA} 
In this appendix we review some necessary group theory for the computations we have done in the text. Let us begin by setting our conventions of $so(1,3)$ algebra and its representations. We take its generators to be $M_{\mu\nu}$ with the algebra:
\be
[M_{\alpha\beta}, M_{\gamma\delta}] = \eta_{\alpha \delta} M_{\beta \gamma} + \eta_{\beta \gamma} M_{\alpha\delta} - \eta_{\alpha \gamma} M_{\beta\delta} - \eta_{\beta\delta} M_{\alpha\gamma}
\ee
where $\mu, \nu, \, \cdots = 0, 1,2,3$ and $\eta = {\rm diag}\{-1,1,1,1\}$. One way of writing this algebra is as $so(1,3) = su(2) \oplus su(2)$ with the generators:
\be
J^{(\pm)}_1 = \frac{1}{2}(-i M_{23} \pm  M_{01}), ~~ J^{(\pm)}_2 = \frac{1}{2}(-i M_{31} \pm M_{02}), ~~ J^{(\pm)}_3 = \frac{1}{2}(-iM_{12} \pm M_{03})
\ee
with the algebra
\be
[J^{(\pm)}_a, J^{(\pm)}_b] = i \, \epsilon_{abc} J^{(\pm)}_c, ~~ [J^{(\pm)}_a, J^{(\mp)}_b] = 0.
\ee
Working with unitary representations for each $SU(2)$ factor provides a finite dimensional non-unitary representation of $so(1,3)$. So a general finite dimensional non-unitary representation is labeled by two half-integers $(j_1, j_2)$. We are interested in constructing representations of the ``diagonal" $SU(2)$ out of these representations labeled by one single $j$. Another way to write the algebra of $so(1,3)$ is as $sl(2, {\mathbb R}) \oplus sl(2, {\mathbb R})$ with the generators:
\be
J^{(\pm)}_1 = \frac{i}{2}(-i M_{23} \pm  M_{01}), ~~ J^{(\pm)}_2 = \frac{i}{2}(-i M_{31} \pm M_{02}), ~~ J^{(\pm)}_0 = \frac{1}{2}(-iM_{12} \pm M_{03})
\ee
%
with the algebra
\be
[J^{(\pm)}_a, J^{(\pm)}_b] =  i \, {\epsilon_{ab}}^c J^{(\pm)}_c, ~~ [J^{(\pm)}_a, J^{(\mp)}_b] = 0.
\ee
with $\epsilon_{012} = 1$ and $\eta_{ab} = {\rm diag} \{-1,1,1\}$ used to raise and lower indices. Working with unitary representations of each $sl(2,{\mathbb R})$ factor provides infinite dimensional but non-unitary representations of $so(1,3)$. These generators of $sl(2, {\mathbb R})$ can be mapped to the standard ones used in the 2d CFT language by defining:
%
%
\bea
L_0 &=& - J^{(+)}_0, ~~ L_1 = i(J^{(+)}_1 + i \, J^{(+)}_2), ~~ , ~~ L_{-1} = -i(J^{(+)}_1 - i \, J^{(+)}_2) \cr
\bar L_0 &=& J^{(-)}_0, ~~ \bar L_{1} =-i( J^{(-)}_1 - i \, J^{(-)}_2), ~~ \bar L_{-1} = i(J^{(-)}_1 + i \, J^{(-)}_2)
\eea
which satisfy the algebra:
\bea
[L_m, L_n] = (m-n) L_{m+n}, ~~ [\bar L_m, \bar L_n] = (m-n) \, \bar L_{m+n}, ~~ [L_m, \bar L_n] = 0.
\eea
The representation where the generators have the hermiticity properties:
%
\bea
L_0^\dagger = L_0, ~ L_1^\dagger = L_{-1}; \bar L_0^\dagger = \bar L_0, ~ \bar L_1^\dagger = \bar L_{-1}
\eea
%
 is the relevant one for us here. We will consider the unitary highest (lowest) weight representation of each of these two $sl(2,{\mathbb R})$ as in, for instance, \cite{Bargmann:1946me, Jackiw:1990ka}. The sub-algebra $so(3)$ in $so(1,3)$ is generated by:
%
\be
L_0 - \bar L_0 = i M_{12}, ~~ L_1 + \bar L_{-1} = i M_{23} +  \, M_{13}, ~~ L_{-1} + \bar  L_1 = -i M_{23} + M_{13}
\ee
The rest of the generators are
\bea
- M_{03} = L_0  + \bar L_0, ~~ M_{01} +i M_{02} = - L_1 + \bar L_{-1}, ~~ M_{01} - i M_{02} =  L_{-1} - \bar L_1
\eea
So the finite dimensional representation of the ``local Lorentz algebra" $so(3)$ are thus associated to the finite dimensional non-unitary representation of the twisted diagonal $sl(2, {\mathbb R})$ generated by $L_n - (-1)^n \bar L_{-n}$ for $n = -1, 0, 1$.

We are interested in decomposing each of the representations of $so(1,3)$ given by the tensor product of the infinite dimensional unitary representation of each of the $sl(2, {\mathbb R})$ algebras in $so(1,3)$ into a given irreducible representation of the twisted diagonal $sl(2, {\mathbb R})$ sub-algebra. 

The fundamental (and the defining representation) of the Lorentz algebra $so(1,3)$ is the vector representation in which we take the generators to be $4\times 4$ real trace-less matrices given by:
\be
{(M_{ab})^\alpha}_\beta = \delta^\alpha_a \, \eta_{b \beta} - \delta^\alpha_b \eta_{a\beta}
\ee
with $\eta^{-1} (-M_{\mu\nu}^{\rm T}) \eta = M_{\mu\nu}$. The $6 \times 6$ adjoint representation is given by
\bea
{(M_{ab})^{gh}}_{mn} = \eta_{an} (\delta^g_b \delta^h_m - \delta^g_m \delta^h_b) + \eta_{bm} (\delta^g_a \delta^h_n - \delta^g_n \delta^h_a) \\ - 
 \eta_{am} (\delta^g_b \delta^h_n - \delta^g_n \delta^h_b) -  \eta_{bn} (\delta^g_a \delta^h_m - \delta^g_m \delta^h_a)
\eea
such that one has
\be
[M_{ab}, M_{mn}] = \frac{1}{2} {(M_{ab})^{gh}}_{mn} M_{gh} = -\frac{1}{2} {(M_{mn})^{gh}}_{ab} M_{gh}
\ee
Defining $O_{[ab][cd]} = \eta_{ac} \eta_{bd} - \eta_{ad} \eta_{bc}$ and the identity matrix as ${I_{[ab]}}^{[cd]} = \delta_a^c \delta_b^d - \delta_a^d \delta_b^c$ we again have $O^{-1} (-M^{\rm T}) O = M$. At the level of the group elements in the adjoint representation this translates to $R[g^{-1}] = O^{-1} (R[g])^{\rm T} O$ where $R[g]$ is the $6 \times 6$ adjoint representation of the group element $g$. 

There are two quadratic Casimirs of the Lie algebra $so(1,3)$:
\begin{itemize}
 \item (i) $C^{(1)}_2 =  M_{0c} M_{0c} - \frac{1}{2} M_{ab} M_{ab}$ 
 \item (ii) $C^{(2)}_2 = M_{01} M_{23} + M_{23} M_{01} - M_{02} M_{13} - M_{13} M_{02} + M_{03} M_{12} + M_{12} M_{03}$. 
 \end{itemize}
Written in terms of the two $sl(2, {\mathbb R})$ factors these read:
\bea
C_2^{(1)} = C_2 + \bar C_2, ~~~ i \, C_2^{(2)} = C_2 - \bar C_2
\eea
or equivalently
\bea
C_2:= 2L_0 L_0 - \{L_1, L_{-1} \} = \frac{1}{2} [ C_2^{(1)} + i \, C_2^{(2)} ], ~
\bar C_2 := 2 \bar L_0 \bar L_0 - \{\bar L_1, \bar L_{-1} \} = \frac{1}{2} [ C_2^{(1)} - i \, C_2^{(2)} ] \nonumber
\eea
\subsection*{\underline{Hermitian representations of $sl(2, {\mathbb R})$}}
Let us review some facts regarding the hermitian representations of the algebra $sl(2, {\mathbb R})$ as in \cite{Jackiw:1990ka}. The quadratic Casimir operator of the algebra $[L_m, L_n] = (m-n) \, L_{m+n}$ is again
\bea
C_2 = 2 L_0 L_0 - 2 L_0 - 2 L_{-1} L_1 = 2 L_0 L_0 + 2 L_0 - 2 L_1 L_{-1}
\eea
 Consider a state $|h, 0 \rangle$ which is a highest weight state: $L_1 | h, 0 \rangle = 0$ and $L_0| h, 0 \rangle = h | h, 0 \rangle $ with $C_2 | h, 0 \rangle = 2 h (h-1) | h, 0 \rangle$. The rest of the states in this representation can be obtained by successively operating with $L_{-1}$ starting from the highest weight state, that is $| h, n \rangle \sim L_{-1}^n |h, 0 \rangle$. Thus we have states in this highest weight representation given by $|h, n\rangle$ such that 
 \bea
 L_0 |h, n \rangle = (h+n) |h, n \rangle, ~~ C_2 |h, n \rangle = 2h (h-1) |h, n\rangle.
 \eea
 which in-turn imply 
 \bea
 L_{-1} |h, n \rangle &=& \sqrt{ (2h+n) (n+1)} ~| h, n+1 \rangle, \cr
 L_1 |h, n \rangle &=& \sqrt{n (2h +n -1)} ~|h, n-1 \rangle.
 \eea
For a positive $h$ there is no state $|h, n \rangle$ with a non-negative integer $n$ which is annihilated by $L_{-1}$ and so they are all infinite dimensional and unitary. 

\subsection*{\underline{Clebsch-Gordan coefficients}}
We need some more group theory - namely, the Clebsch-Gordan coefficients that appear in the decomposition of the tensor product of two unitary representations of $sl(2, {\mathbb R})$ algebra into other unitary irreducible ones. We take the basis states of the tensor product of two irreducible representations of $sl(2, {\mathbb R})$ labeled by (the non-negative real numbers) $h_1$ and $h_2$ by $|h_1, n_1 \rangle \otimes |h_2, n_2 \rangle = | h_1, h_2; n_1, n_2 \rangle$ which diagonalize $\{C_2^{(1)}, C_2^{(2)}, L^{(1)}_0, L^{(2)}_0\}$  with the eigen values $\{ 2h_1(h_1-1), 2h_2(h_2-1), h_1+n_1, h_2+n_2 \}$ respectively. The generators of $sl(2, {\mathbb R})$ act on the tensor product as
\bea
L_n = L^{(1)}_n \otimes {\mathbb I} + {\mathbb I} \otimes L^{(2)}_n \, .
\eea
We make a change of basis to states $| h_1, h_2; h, n \rangle$ which diagonalize $\{C_2^{(1)}, C_2^{(2)}, C_2, L_0 \}$ with eigen values $\{ 2h_1(h_1-1), 2h_2(h_2-1), 2h(h-1), h+n \}$ respectively. We also have
 \bea
 L_0 |h_1, h_2; h, n \rangle &=& (h+n) |h_1, h_2; h, n \rangle \cr
 L_{-1} |h_1, h_2; h, n \rangle &=& \sqrt{ (2h+n) (n+1)} ~|h_1, h_2; h, n+1 \rangle, \cr
 L_1 |h_1, h_2; h, n \rangle &=& \sqrt{n (2h +n -1)} ~|h_1, h_2; h, n-1 \rangle.
 \eea
Taking the matrix element of the operator $L_0 - L_0^{(1)} - L_0^{(2)}$ as
\bea
\langle h_1, h_2; h, n  |L_0 - L_0^{(1)} - L_0^{(2)}| h_1, h_2; n_1, n_2 \rangle \\ = (h+n - h_1-n_1-h_2-n_2) \langle h_1, h_2; h, n  | h_1, h_2; n_1, n_2 \rangle = 0
\eea
Imposes the condition 
\bea
h- h_1 - h_2 = n_1+n_2 - n.
\eea
Given that $n_1, n_2, n$ are integer the allowed values of $h$ are discrete and integer spaced.  %
On the other hand considering the action of $L_1$ gives:
 \bea
&&\sqrt{n (2h +n -1)} ~\langle h_1, h_2; n_1, n_2 |h_1, h_2; h, n-1 \rangle \cr
&& ~~~~~~ = ~~~~ \sqrt{(n_1+1) (2h_1 +n_1)} ~ \langle h_1, h_2; n_1+1, n_2| h_1, h_2; h, n \rangle \cr
&& ~~~~~~~~~~~ + \sqrt{(n_2+1) (2h_2 +n_2)} ~\langle h_1, h_2; n_1, n_2+1| h_1, h_2; h, n \rangle 
\eea
and the action of $L_{-1}$ gives:
 \bea
&&\sqrt{(n+1) (2h +n)} ~\langle h_1, h_2; n_1, n_2 |h_1, h_2; h, n +1 \rangle \cr
&& ~~~~~~ = ~~~~ \sqrt{n_1 (2h_1 +n_1-1)} ~ \langle h_1, h_2; n_1-1, n_2| h_1, h_2; h, n \rangle \cr
&& ~~~~~~~~~~ + \sqrt{n_2 (2h_2 +n_2 -1 )} ~\langle h_1, h_2; n_1, n_2-1| h_1, h_2; h, n \rangle 
\eea
when we write 
\bea
| h_1, h_2; h, n \rangle = \sum_{m_1, m_2 = 0}^\infty |h_1, h_2, m_1, m_2 \rangle \langle h_1, h_2; m_1, m_2| h_1, h_2; h, n \rangle
\eea
and when $|h_1, h_2, m_1, m_2 \rangle$ and $| h_1, h_2; h, n \rangle$ are basis vectors of unitary representations (as it is when $h_1, h_2, h > 0$) one should impose
\bea
\sum_{h, n} \langle h_1, h_2; n_1, n_2 | h_1, h_2; h, n \rangle \langle h_1, h_2; m_1, m_2 | h_1, h_2; h, n \rangle &=& \delta_{n_1, m_1} \delta_{n_2, m_2}
\eea
Later on we will need explicit expressions for these CG coefficients. So let us derive them here (see also \cite{holman}). To proceed let us first write the CG coefficient $\langle h_1, h_2; n_1, n_2 |h_1, h_2; h, n \rangle $ as 
\bea
\langle h_1, h_2; n_1, n_2 |h_1, h_2; h_3, n_3 \rangle = \frac{1}{\prod_{i=1}^3 \sqrt{k_i!(\Gamma(2h_i+k_i)}} f(k_1, k_2;k_3)
\eea
where the f's satisfy
\bea
\label{recrln2}
k_3 (2h_3+k_3 - 1) f(k_3 -1) - f(k_1+1) - f(k_2+1) &=& 0 \\
\label{recrln3}
f(k_3 + 1) - k_1 (2h_1 + k_1 - 1) f(k_1-1) - k_2 (2h_2 + k_2 - 1) f(k_2-1) &=& 0 
\eea
modulo $h_1+k_1 + h_2+k_2 - h_3-k_3$ (as a consequence of the Kronecker delta in the CG coefficient). There are two cases we have to consider carefully: $h_1+h_2 -h_3 \ge 0$ and $h_1+h_2 - h_3 \le 0$. The CG coefficients are:
\bea
\label{ourcgs}
f(k_1, k_2; k_3) & \sim &  \Gamma(2h_2+k_1+k_2) \Gamma(k_3-k_2) k_2! ~ \delta_{h_1+k_1+h_2+k_2-h_3-k_3} \cr  && \times {}_3F_2 \left( {{-k_1, -k_3, h_3 + h_1 -h_2 } \atop {1+k_2 - k_3, 1-2h_2 - k_1 - k_2}};1\right) ~~~ {\rm for} ~~~ h_i+h_j \ge h_k ~~ {\rm or} ~~  h_3+h_1 \le h_2\cr
&\sim & (-1)^{k_1}  \tfrac{\Gamma(2h_2+k_1+k_2)  k_2!}{\Gamma(k_2-k_3+1)} ~\delta_{h_1+k_1+h_2+k_2-h_3-k_3} \cr  && \times {}_3F_2 \left( {{-k_1, -k_3, h_3 + h_1 -h_2 } \atop {1+k_2 - k_3, 1-2h_2 - k_1 - k_2}};1\right)  ~~~ {\rm for} ~~~ h_1+h_2 \le h_3 \cr
&\sim & (-1)^{k_3} \tfrac{ \Gamma(k_3-k_2) k_2!}{\Gamma(h_1-h_2-h_3+k_3+1)} ~\delta_{h_1+k_1+h_2+k_2-h_3-k_3} \cr
&& \times {}_3F_2 \left( {{-k_1, -k_3, h_3 + h_1 -h_2 } \atop {1+k_2 - k_3, 1-2h_2 - k_1 - k_2}};1\right)  ~~~ {\rm for} ~~~ h_2+h_3 \le h_1
\eea
Substituting them (any of the above four cases) into the left hand side of (\ref{recrln2}) ((\ref{recrln3})) comes out to be proportional to $\delta_{h_1+k_1+h_2+k_2-h_3-k_3+1}$ ($\delta_{h_1+k_1+h_2+k_2-h_3-k_3-1}$) times
\bea
\label{hgidentlhs}
d e {}_3F_2 \left( {{a, b, c } \atop {d, e}};1\right) + a (c-d) {}_3F_2 \left( {{a+1, b+1, c } \atop {d+1, e+1}};1\right) + d (a-e) {}_3F_2 \left( {{a, b+1, c } \atop {d, e+1}};1\right) \nonumber
\eea
where $a=-k_3$, $b=-k_1-1$, $c=h_3+h_1-h_2$, $d=-2h_2-k_1-k_2$ and $e=1+k_2-k_3$ ($a=-k_1$, $b= -k_3-1$, $c=h_3+h_1-h_2=k_2-k_3-k_1-1$, $d=k_2-k_3$ and $e = 1-2h_2-k_1-k_2$) for either case $h_1+h_2 \ge h_3$ or $h_1+h_2 \le h_3$. Happily this vanishes identically (at least for when either $a$ or $b$ is a negative integer as it is the case in our case). The interested reader may consult appendix \ref{appendixd} for a proof of this identity. 

Next we will establish two important relations satisfied by the CG coefficients of $sl(2, {\mathbb R})$ and the unitary representations of the algebra elements. First of them is
\bea
\label{CGidentity1}
(L_n^{(h_1)})_{k_1 k_1'} C^{h_1h_2;h}_{k_1'k_2k_3} + (L_n^{(h_2)})_{k_2 k_2'} C^{h_1h_2;h}_{k_1k_2'k_3} -  C^{h_1h_2;h}_{k_1k_2k_3'} (L_n^{(h_3)})_{k_3' k_3}= 0
\eea
where $n=-1, 0, 1$. To verify this first note that the matrix elements of the generators $L_n$ are:
\bea
(L_0^{(h)})_{k, k'} &=& (h+k) \delta_{k,k'}, \cr
 (L_1^{(h)})_{k,k'} &=& \sqrt{(k+1) (2h+k)} ~ \delta_{k, k'-1}, \cr
  (L_{-1}^{(h)})_{k,k'} &=& \sqrt{k (2h+k-1)} ~ \delta_{k, k'+1}
\eea
Taking the case of $n=0$ implies that the CG coefficients vanish unless $h_1+k_1+h_2+k_2= h+k$. It is easily seen that for $n=-1,1$ this identity follows simply from the recursion relations satisfied by the CG coefficients. The second identity is
\bea
\label{noxatvertex}
R_{h_1}[g(x)]_{k_1k_1'} ~ R_{h_2}[g(x)]_{k_2,k_2'} ~ C^{h_1h_2;h}_{k_1'k_2';k_3'} ~R_{h_3}[g^{-1}(x)]_{k_3'k_3} = C^{h_1h_2;h}_{k_1k_2;k_3}
\eea
where $R_h[g(x)]$ is the representation of the group element $g(x)$ in the lowest weight representation of $sl(2,{\mathbb R})$ labeled by $h$. We will now prove it for an element of the $SL(2, {\mathbb R})$ group of the form $g(x) = e^{\omega^a(x) L_a}$ (summed over $a = -1,0,1$). Then we can write $R_h[g(x)] = \sum_{n=0}^\infty \frac{1}{n!} \omega^a (L_a^{(h)})^n$ where $L_a^{(h)}$ is the representation of the generator $L_a$ in representation labeled by $h$. To establish this identity we look at terms involving a fixed number of parameters $\omega^a$. To the order ${\cal O}(\omega_a^0)$ the right hand side is already taken care of. At the ${\cal O}(\omega_a^1)$ the terms sum to zero from (\ref{CGidentity1}). At the ${\cal O}(\omega_a^2)$ we have
\bea
&& \omega^a \omega^b  \big[ \frac{1}{2} (L_a^{(1)}L_b^{(1)})_{k_1k_1'} C(k_1') +  \frac{1}{2} (L_a^{(2)}L_b^{(2)})_{k_2k_2'} C(k_2') +  \frac{1}{2} C(k_3')  (L_a^{(3)}L_b^{(3)})_{k_3'k_3} \cr
&&+ (L^{(1)}_a)_{k_1k_1'} (L^{(2)}_b)_{k_2k_2'} C(k_1', k_2') - (L^{(1)}_a)_{k_1k_1'}  C(k_1', k_3') (L^{(3)}_b)_{k_3'k_3}- (L^{(2)}_a)_{k_2k_2'}  C(k_2', k_3') (L^{(3)}_b)_{k_3'k_3} \big] \nonumber \cr
&=& \frac{1}{2} \omega^a (L_a^{(1)})_{k_1k_1''} \, \omega^b \big[ (L_b^{(1)})_{k_1''k_1'} C(k_1') + (L_b^{(2)})_{k_2 k'_2} C(k_1'', k_2') - C(k_1'',k_3') (L_b^{(3)})_{k_3'k_3}\big] \cr
&& + \frac{1}{2} \omega^a (L_a^{(2)})_{k_2k_2''} \, \omega^b \big[ (L_b^{(2)})_{k''_2k'_2} C(k'_2) + (L_b^{(1)})_{k_1 k'_1} C(k'_1, k''_2) - C(k''_2,k_3') (L_b^{(3)})_{k_3'k_3}\big] \cr
&& + \frac{1}{2}  \omega^b \big[ C(k'_3) (L_b^{(3)})_{k'_3 k''_3}  - (L_b^{(1)})_{k_1 k'_1} C(k'_1, k''_3)  - (L_b^{(2)})_{k_2 k'_2} C(k'_2, k''_3)\big] \omega^a (L_a^{(3)})_{k''_3k_3} \,
\eea
And using the identity (\ref{CGidentity1}) each of the three terms vanishes. One can generalise this to any higher order in powers of $\omega_a$'s establishing the identity we claimed. The term of ${\cal O}(\omega^{n+1})$ is
\bea
&& \sum_{n_1+n_2+n_3 = n+1} \frac{(-1)^{n_3}}{n_1! n_2! n_3!} [(\omega^a L_a^{(1)})^{n_1}]_{k_1k'_1} [(\omega^b L_b^{(2)})^{n_2}]_{k_2k'_2} C(k'_1, k'_2; k'_3) [(\omega^c L_c^{(3)})^{n_3}]_{k'_3k_3} \cr
&=& \frac{1}{n+1} \omega^d (L_d^{(1)})_{k_1 k'_1} \sum_{n_1+n_2+n_3 = n} \frac{(-1)^{n_3}}{n_1! n_2! n_3!} [(\omega^a L_a^{(1)})^{n_1}]_{k'_1k''_1} [(\omega^b L_b^{(2)})^{n_2}]_{k_2k'_2} C(k''_1, k'_2; k'_3) [(\omega^c L_c^{(3)})^{n_3}]_{k'_3k_3} \cr
&+&  \frac{1}{n+1} \omega^d (L_d^{(2)})_{k_2 k'_2} \sum_{n_1+n_2+n_3 = n} \frac{(-1)^{n_3}}{n_1! n_2! n_3!} [(\omega^a L_a^{(1)})^{n_1}]_{k_1k'_1} [(\omega^b L_b^{(2)})^{n_2}]_{k'_2k''_2} C(k'_1, k''_2; k'_3) [(\omega^c L_c^{(3)})^{n_3}]_{k'_3k_3} \cr
&-& \frac{1}{n+1}  \omega^d (L_d^{(3)})_{k'_3k_3}  \sum_{n_1+n_2+n_3 = n} \frac{(-1)^{n_3}}{n_1! n_2! n_3!} [(\omega^a L_a^{(1)})^{n_1}]_{k_1k'_1} [(\omega^b L_b^{(2)})^{n_2}]_{k_2k'_2} C(k'_1, k'_2; k''_3) [(\omega^c L_c^{(3)})^{n_3}]_{k''_3k'_3}  \nonumber \cr
\eea
Therefore assuming the relation to be true at $n^{th}$ order means it is true at order $n+1^{th}$ order. This establishes the identity we want.

\section{Proof of eq.(\ref{kvfromadjg})}
\label{appendixb}
 For this, let us define the vector operator (an element of the Lie algebra):
\bea
\xi_\mu = e^a_\mu \, g^{-1} \! M_{0a} \, g.
\eea
and calculate $D_\mu \xi_\nu = \partial_\mu \xi_\nu - \Gamma^\lambda_{\mu\nu} \xi_\lambda$. One can show using the equations satisfied by $g$ and $g^{-1}$ (\ref{gequation}, \ref{igequation}) that
\bea
\partial_\mu \xi_\nu = (\partial_\mu e^a_\nu + {\omega_\mu}^{ac} e^c_\nu) \, g^{-1} M_{0a} g + \frac{1}{l} e^a_\mu e^b_\nu \, g^{-1} \! M_{ab} g = \Gamma^\lambda_{\mu\nu} e^a_\lambda \, g^{-1} \! M_{0a} g + \frac{1}{l} e^a_\mu e^b_\nu \, g^{-1} \! M_{ab} g \nonumber \\
\eea
where we have used $\partial_\mu e^a_\nu + {\omega_\mu}^{ac} e^c_\nu - \Gamma^\lambda_{\mu\nu} e^a_\lambda = 0$. This implies
\bea
D_\mu \xi_\nu = \frac{1}{l} e^a_\mu e^b_\nu \, g^{-1} \! M_{ab} g \, .
\eea
Taking the symmetric part clearly shows that the vector operators $\xi_\mu$ satisfy the Killing vector equations $D_\mu \xi_\nu + D_\nu \xi_\mu  = 0$. This result generalises how one obtains Killing vectors from bilinear of Killing spinors to arbitrary representations of the local lorentz group. This Killing vector operator can be expanded into a linear combination of the generators
\bea
\xi_\mu = e^a_\mu \left[ {(R[g^{-1}])^{0b}}_{0a} M_{0b} + \frac{1}{2}  {(R[g^{-1}])^{bc}}_{0a} M_{bc} \right]
\eea
where $R[g^{-1}]$ is the representation of $g^{-1}$ in the adjoint representation. One can raise the index on the Killing operator as
$\xi^\mu = E_a^\mu g^{-1} M_{0a} g$:
\bea
\xi^\mu = E_a^\mu \left[ {(R[g^{-1}])^{0b}}_{0a} M_{0b} + \frac{1}{2}  {(R[g^{-1}])^{bc}}_{0a} M_{bc} \right].
\eea
This enable one to immediately read out the components of the Killing vectors 
\bea
(l^{[\alpha\beta]})^\mu = l \, E_a^\mu {(R[g^{-1}])^{\alpha\beta}}_{0a} 
\eea
or equivalently 
\bea
(l_{[\alpha\beta]})^\mu = - l \, E_a^\mu {(R[g^{-1}])_{\alpha\beta}}^{0a} = - l \, E_a^\mu {(R[g])^{0a}}_{\alpha\beta}
\eea
\section{3-point function - details}
\label{appendixc}
The answer for the three point function of three primary operators in a CFT is
\bea
\label{3ptfnans}
&& \langle O^{(h_1, \bar h_1)}(z_1, \bar z_1) O^{(h_2,\bar h_2)}(z_2, \bar z_2) O^{(h_3, \bar h_3)}(z_3, \bar z_3) \cr  && = \frac{C_{123}}{z_{12}^{h_1+h_2 - h_3} z_{23}^{h_2+h_3 - h_1} z_{13}^{h_3+h_1 - h_2} \bar z_{12}^{\bar h_1+ \bar h_2 - \bar h_3} \bar z_{23}^{\bar h_2+ \bar h_3 - \bar h_1} \bar z_{13}^{\bar h_3+ \bar h_1 - \bar h_2}}
\eea
Without loss of generality we assume $h_3 \ge h_1, h_2$. Then it is clear that $h_3 + h_1 - h_2 \ge 0$ and $h_3 + h_2 - h_1 \ge 0$ with $h_1 + h_2 - h_3$ taking either sign. We consider $h_1 + h_2 - h_3 \ge 0 $ case first -- which we further assume to be an integer $k \in {\mathbb Z}_{\ge 0}$. Let us expand the holomorphic part of the right hand side of (\ref{3ptfnans}) as follows:
\bea
\label{hol3ptfn}
z_2^{h_3-h_1-h_2} z_3^{-2h_3}\sum_{n_i=0}^\infty \tfrac{\Gamma(h_1+h_2 - h_3 +n_3)}{\Gamma(h_1+h_2-h_3)} \tfrac{\Gamma(h_2+h_3-h_1+n_1)}{\Gamma(h_2+h_3-h_1)} \tfrac{\Gamma(h_3+h_1-h_2+n_2)}{\Gamma(h_3+h_1-h_2)} 
\tfrac{z_1^{n_2+n_3}}{n_1 !} \tfrac{z_2^{n_1-n_3}}{n_2 !}\tfrac{z_3^{-n_1-n_2}}{n_3!}
\eea
up to the phase factor $(-1)^{h_1 + h_2+h_3}$. We want to recover this from the computation of our spin-network. 
\bea
&& \sum_{k_i = 0}^\infty \left[ (-1)^{-\frac{h_1}{2}} e^{-\rho h_1} z_1^{k_1} \sqrt{\tfrac{\Gamma(2h_1+k_1)}{k_1! \Gamma(2h_1)}} \right]   \times \left[(-1)^{-\frac{h_2}{2}} e^{-\rho h_2} z_2^{k_2} \sqrt{\tfrac{\Gamma(2h_2+k_2)}{k_2! \Gamma(2h_2)}} \right] \cr
&& \times  \left[ (-1)^{\frac{h_3}{2}} e^{-\rho h_3} z_3^{-2h_3-k_3} \sqrt{\tfrac{\Gamma(2h_3+k_3)}{k_3! \Gamma(2h_3)}} \right]\times \cr
&& \times \tfrac{\Gamma(k_3-k_2) k_2!\Gamma(2h_2+k_1+k_2) \Gamma(h_3+h_1-h_2)}{\sqrt{k_1! k_2! k_3! \Gamma(2h_1+k_1) \Gamma(2h_2 + k_2)\Gamma(2h_3 + k_3)}}  {}_3F_2 \left( {{-k_1, -k_3, 2h_1 + k_1 +k_2 - k_3} \atop {1+k_2 - k_3, 1-2h_2 - k_1 - k_2}};1\right) \delta_{k_1 + k_2 - k_3 + h_1 +h_2 - h_3} \cr
&\sim & z_3^{-2h_3}\sum_{k_i=0}^\infty  \tfrac{1}{k_1! k_3!} \tfrac{\Gamma(2h_2 + k_1 + k_2) \Gamma(k_3- k_2)}{\Gamma(h_1 +h_2 - h_3) \Gamma(2h_2 + k_1 + k_2 - k_3)} {}_3F_2 \left( {{-k_1, -k_3, 2h_1 + k_1 +k_2 - k_3} \atop {1+k_2 - k_3, 1-2h_2 - k_1 - k_2}};1\right) \cr
&& ~~~~~~~~~~~~~~~~~~~~~~ \times z_1^{k_1} 
z_2^{k_2} z_3^{-k_3} \delta_{k_1 + k_2 - k_3 + h_1 +h_2 - h_3} \cr
&\sim & z_3^{-2h_3}\sum_{k_i=0}^\infty \sum_{n=0}^{{\rm min}(k_1, k_3)} \tfrac{\Gamma(k_3 - k_2 - n) \Gamma(2h_2 +k_1 +k_2 - n) \Gamma(2h_1 +k_1 +k_2 - k_3 +n)}{\Gamma(h_1+h_2-h_3) \Gamma(2h_2 + k_1 +k_2 - k_3) \Gamma(2h_1 + k_1 +k_2 - k_3)}  \tfrac{z_1^{k_1} z_2^{k_2} z_3^{-k_3}}{n! (k_1-n)!(k_3-n)!} \, \delta_{k_1 + k_2 - k_3 + h_1 +h_2 - h_3}\cr &&
\eea
where we have used the explicit series representation of ${}_3F_2$. Next we write $k_1 = n_2+n_3$, $k_3 = n_1+n_2$ $n= n_2$ and $k_2 = n_1 - n_3 + h_3-h_1 -h_2 $. Then the four sums can be reduced to three sums over $n_1, n_2, n_3$ at the expense of removing the Kronecker delta to obtain an expression exactly proportional to (\ref{hol3ptfn}). The anti-holomorphic part can be dealt with similarly. The case of $h_1 + h_2 \le h_3$ can also be considered and one can show that the correct functional form of the three-point function can be recovered by using the suitable CG coefficient expression.
\section{A proof that (\ref{hgidentlhs}) vanishes}
\label{appendixd}
We shall now prove the identity
\bea
\label{d1}
de \ _3F_2 \left( {{a, b,c } \atop {d,e}};1\right)&=& a(d-c)\ _3F_2 \left( {{a+1, b+1,c } \atop {d+1,e+1}};1\right)+d(e-a)\ _3F_2 \left( {{a, b+1,c } \atop {d,e+1}};1\right)
\eea
We begin by applying the identity \cite{wolfram}
\bea
a(b-c)\ _3F_2 \left( {{a+1, a_2,a_3 } \atop {b+1,c+1}};z\right)-c(b-a)\ _3F_2 \left( {{a,a_2,a_3 } \atop {b+1,c}};z\right)+b(c-a)\ _3F_2 \left( {{a,a_2,a_3 } \atop {b,c+1}};z\right)=0 \nonumber
\eea
at $z=1$\footnote{we note here that in general the hypergeometric function $_3F_2$ is not well defined at $z=1$, but we will be using it only for the case where  $a=-k_3\pm 1 $, $b=-k_1\pm 1$, $c=h_3+h_1-h_2\pm 1$, $d=-2h_2-k_1-k_2\pm 1$ and $e=1\pm k_2-k_3$ in which case $_3F_{2}$ is a polynomial and the limit is easily taken.} to the right hand side of (\ref{d1}) to give
\bea
a(d-c)\left[\tfrac{e(d-a)}{a(d-e)}\ _3F_2 \left( {{a,b+1,c } \atop {d+1,e}};1\right)-\tfrac{d(e-a)}{a(d-e)}\ _3F_2 \left( {{a,b+1,c } \atop {d,e+1}};1\right)\right] +d(e-a)\ _3F_2 \left( {{a,b+1,c } \atop {d,e+1}};1\right)\nonumber
\eea
which further simplifies to
\bea
\label{d2}
\frac{1}{a(d-e)} \left[ ae(d-a)(d-c)\ _3F_2 \left( {{a,b+1,c } \atop {d+1,e}};1\right)-ad(e-a)(e-c)\ _3F_2 \left( {{a,b+1,c } \atop {d,e+1}};1\right) \right] 
\eea
Now we apply the identity \cite{wolfram}
\bea
b \ _3F_2 \left( {{a, a_2,a_3 } \atop {b,b_2}};z\right)-a \ _3F_2 \left( {{a+1, a_2,a_3 } \atop {b+1,b_2}};z\right)+(a-b)\ _3F_2 \left( {{a, a_2,a_3 } \atop {b+1,b_2}};z\right)=0 \nonumber
\eea
at $z=1$ to both terms of (\ref{d2}) to yield
\bea
\label{d3}
\tfrac{de\left[(d-a)(d-c)-(e-a)(e-c)\right]}{b(d-e)}\ _3F_2 \left( {{a, b,c } \atop {d,e}};1\right)&-&\tfrac{e(d-a)(d-c)(d-b)}{b(d-e)}\ _3F_2 \left( {{a, b,c } \atop {d+1,e}};1\right) \cr
&+& \tfrac{d(e-a)(e-c)(e-b)}{b(e-d)}\ _3F_2 \left( {{a, b,c } \atop {d,e+1}};1\right) 
\eea
Finally we apply the identity \cite{wolfram}
\bea
(a+(a_2+a_3-d-e)z)\ _3F_2 \left( {{a, a_2,a_3 } \atop {d,e}};z\right)+\tfrac{(d-a)(d-a_2)(d-a_3)}{d(d-e)}\ _3F_2 \left( {{a, a_2,a_3 } \atop {d+1,e}};z\right)+\cr \tfrac{(e-a)(e-a_2)(e-a_3)}{e(e-d)}\ _3F_2 \left( {{a, a_2,a_3 } \atop {d,e+1}};z\right)-a(1-z)\ _3F_2 \left( {{a+1, a_2,a_3 } \atop {d,e}};z\right)=0\nonumber
\eea
to (\ref{d3}) at $z=1$ to get 
\bea
\tfrac{de}{b} \left[(d+e-a-c)+(a+b+c-d-e) \right] \ _3F_2 \left( {{a,b,c } \atop {d,e}};1\right) \nonumber
=de\ _3F_2 \left( {{a,b,c } \atop {d,e}};1\right)\nonumber
\eea
which proves our claim.

\bibliographystyle{utphys}
\providecommand{\href}[2]{#2}\begingroup\raggedright\endgroup
\end{document}